\documentclass[prb,aps]{revtex4}
\usepackage{graphicx}

\begin{document}
\title{\Large Anomalous scaling of spin accumulation in ferromagnetic tunnel devices with silicon and germanium}
\author{S. Sharma$^{1,2}$, A. Spiesser$^{1}$, S. P. Dash$^{3}$, S. Iba$^{1}$, S. Watanabe$^{1,4}$, B. J. van Wees$^{2}$, H. Saito$^{1}$, S. Yuasa$^{1}$ and R. Jansen$^{1}$}
\affiliation{$^1$\,National Institute of Advanced Industrial
Science and Technology (AIST), Spintronics Research Center, Tsukuba, Ibaraki 305-8568, Japan.\\
$^2$\,Zernike Institute for Advanced Materials, Physics of
Nanodevices, University of Groningen, 9747 AG, Groningen, The Netherlands.\\
$^3$\,Department of Microtechnology and Nanoscience, Chalmers
University of Technology, SE-41296, G\"{o}teborg, Sweden.\\
$^4$\,University of Tsukuba, Tsukuba, Ibaraki 305-8571, Japan}


\begin{abstract}
The magnitude of spin accumulation created in semiconductors by electrical
injection of spin-polarized electrons from a ferromagnetic tunnel
contact is investigated, focusing on how the spin signal detected
in a Hanle measurement varies with the thickness of the tunnel oxide.
An extensive set of spin-transport data for Si and Ge magnetic tunnel
devices reveals a scaling with the tunnel resistance that violates the
core feature of available theories, namely, the linear proportionality
of the spin voltage to the injected spin current density. Instead, an
anomalous scaling of the spin signal with the tunnel resistance is observed,
following a power-law with an exponent between 0.75 and 1 over 6 decades.
The scaling extends to tunnel resistance values larger than 10$^{9}$ $\Omega\mu m^2$,
far beyond the regime where the classical impedance mismatch
plays a role. This scaling is incompatible with existing theory
for direct tunnel injection of spins into the semiconductor. It also
demonstrates conclusively that the large spin signal does not originate from
two-step tunneling via localized states near the oxide/semiconductor interface.
Control experiments on devices with a non-magnetic metal (Ru) electrode,
instead of the semiconductor, exhibit no Hanle spin signal, showing that
spin accumulation in localized states within the tunnel barrier is
also not responsible. Control devices in which the spin current is removed
by inserting a non-magnetic interlayer exhibit no Hanle signals either,
proving that the spin signals observed in the standard devices are genuine
and originate from spin-polarized tunneling and the resulting spin accumulation.
Altogether, the scaling results suggest that the spin signal is proportional
to the applied bias voltage, rather than the (spin) current.
\end{abstract}

\maketitle

\section{Introduction}
\indent Mainstream semiconductors such as silicon and germanium play a key role in the
development of a spintronics information technology in which spin is used to represent
digital data \cite{chappert,awschalom,fertnobel,jansennmatreview}. To create and detect
spin-polarized carriers in non-magnetic materials, the use of ferromagnetic tunnel
contacts has proven to be a robust and technologically viable approach that is widely
used in spin-based devices, including those with Si and Ge
\cite{jonker,erve,dash,jansencs,sasakitdep,suzuki,jeon,hamayasi3,hamayaefield,toshiba1,toshiba2,saitoge,wangge,jeonge,ibage,ibagert,jain,hamayaschottky}.
As recently reviewed \cite{jansennmatreview,jansensstreview}, controversy has arisen
because in many semiconductor spintronic devices, the magnitude of the observed spin
voltage differs by several orders of magnitude from what is expected based on the
available theory for spin injection and diffusion \cite{fertPRB,fertIEEE,maekawa,dery}.
A common feature of all theories is that the injected spin current produces a spin accumulation
$\Delta\mu$, i.e., a spin splitting in the electrochemical potential and thus a spin-dependent
occupation of the electronic states in the non-magnetic material. Conservation of spin-angular
momentum requires the injected spin current to be balanced by spin relaxation, from which the
steady-state non-equilibrium spin accumulation is evaluated. Consequently, the spin accumulation
is predicted to be linearly proportional to the injected spin current.\\
\indent A powerful way to test the predictions is to vary the thickness of the tunnel barrier,
which changes the current density $J$ exponentially. The spin accumulation is expected to
exhibit a similar exponential variation, so that $\Delta\mu / J$ remains constant. Here we
present an extensive set of spin-transport data on Si and Ge based magnetic tunnel devices
with different tunnel oxides. The scaling of the detected spin voltage with tunnel oxide thickness
violates the expected linear proportionality of spin voltage and injected spin current. The data
is shown to be incompatible with any of the known theories, including those based on direct tunneling
\cite{fertPRB,fertIEEE,maekawa,dery} or two-step tunneling via localized states
\cite{tran,jansentwostep}.

\section{Device fabrication}
\indent To illustrate the generic nature of the observed scaling, we use devices with
heavily doped Si (p-type and n-type) as well as p-type Ge, with an amorphous Al$_2$O$_3$
tunnel barrier and Ni$_{80}$Fe$_{20}$ ferromagnet, or with epitaxial, crystalline MgO/Fe
contacts. Tunnel contacts on Si(001) surfaces were fabricated using $n$-type
silicon-on-insulator wafers with a 5 $\mu$m thick active Si layer having As-doping
and a resistivity of 3 m$\Omega$cm at 300 K, or $p$-type silicon-on-insulator
wafers with a 3 $\mu$m thick active Si layer having B-doping and a resistivity of
11 m$\Omega$cm at 300 K. For the contacts with amorphous Al$_{2}$O$_{3}$, the Si
substrate was treated by hydrofluoric acid to remove oxide, the substrate was
introduced into the ultrahigh vacuum chamber, and the tunnel barrier was prepared
by electron-beam deposition of Al$_{2}$O$_{3}$ from a single-crystal
Al$_{2}$O$_{3}$ source, followed by plasma oxidation for 2.5 minutes, and
electron-beam deposition of the ferromagnetic-metal top electrode (typically
10 nm thick) and a Au cap layer, all at room temperature. For some devices
(appendix A) the plasma oxidation step was omitted. The plasma
oxidation leads to the formation of some additional silicon oxide. The actual
tunnel oxide thickness is thus slightly larger than the nominal thickness of the
deposited Al$_{2}$O$_{3}$, as previously confirmed and quantified \cite{minthesis}
by transmission electron microscopy (TEM). Values of the tunnel oxide thickness
quoted in this manuscript correspond to the corrected, actual oxide thickness
extracted from TEM, and for convenience this is referred to as the Al$_{2}$O$_{3}$
thickness.\\
\indent For epitaxial contacts \cite{spiesserspie} with MgO/Fe, after treatment with buffered hydrofluoric
acid, the Si substrate was annealed in the ultrahigh vacuum deposition system
to 700$^{\circ}$C for 10 minutes to obtain a 2$\times$1 reconstructed Si
surface. The MgO tunnel barrier and the Fe electrode (10 nm thick) were deposited
at 300$^{\circ}$C and 100$^{\circ}$C, respectively, and the crystalline
nature of the layers was confirmed by {\em in situ} reflection high-energy
electron diffraction and by high-resolution TEM, as reported recently \cite{spiesserspie}.
Tunnel devices on p-type Ge(001) were prepared using Ga-doped wafers with a resistivity of
3 m$\Omega$cm and a carrier concentration of 8.2$\times$10$^{18}$ cm$^{-3}$ at 300 K. The preparation of the
epitaxial MgO/Fe tunnel contacts on Ge was as previously described \cite{ibagert}.\\
\indent To probe the spin accumulation over a large range of the tunnel barrier thickness,
we employ three-terminal devices \cite{dash} in which a single ferromagnetic tunnel contact
is used to inject the spin accumulation, and to detect it. This geometry, unlike 4-terminal non-local
devices \cite{erve,suzuki}, allows the contact area to be chosen arbitrarily
large so as to adjust the overall device resistance and thereby ensure a sufficient signal
to noise ratio. Here, the tunnel junction area is between 10$\times$10 $\mu m^2$ and
100$\times$200 $\mu m^2$. Positive voltage corresponds to electrons tunneling from ferromagnet to semiconductor.

\section{Scaling with tunnel barrier thickness}
\indent In all devices, voltage signals corresponding to the Hanle and inverted Hanle
effect \cite{invertedhanle} were detected when a magnetic field is applied perpendicular or parallel to
the tunnel interface, respectively, at constant tunnel current (Fig. 1). The Hanle (inverted Hanle)
signal originates from the suppression (recovery) of the spin accumulation due to spin precession
(or the reduction thereof), and is the signature of the presence of a spin accumulation
\cite{dash,invertedhanle}. The most striking observation is that the amplitude of
the spin signal (the spin RA product, defined as $\Delta V_{Hanle}/J$, the spin voltage
signal per unit of $J$) increases by orders of magnitude when the thickness of
the tunnel barrier is increased. The width of the Hanle curve and the ratio of the Hanle and
inverted Hanle amplitudes are also not constant.

\begin{figure}[htb]
\hspace*{0mm}\includegraphics*[width=72mm]{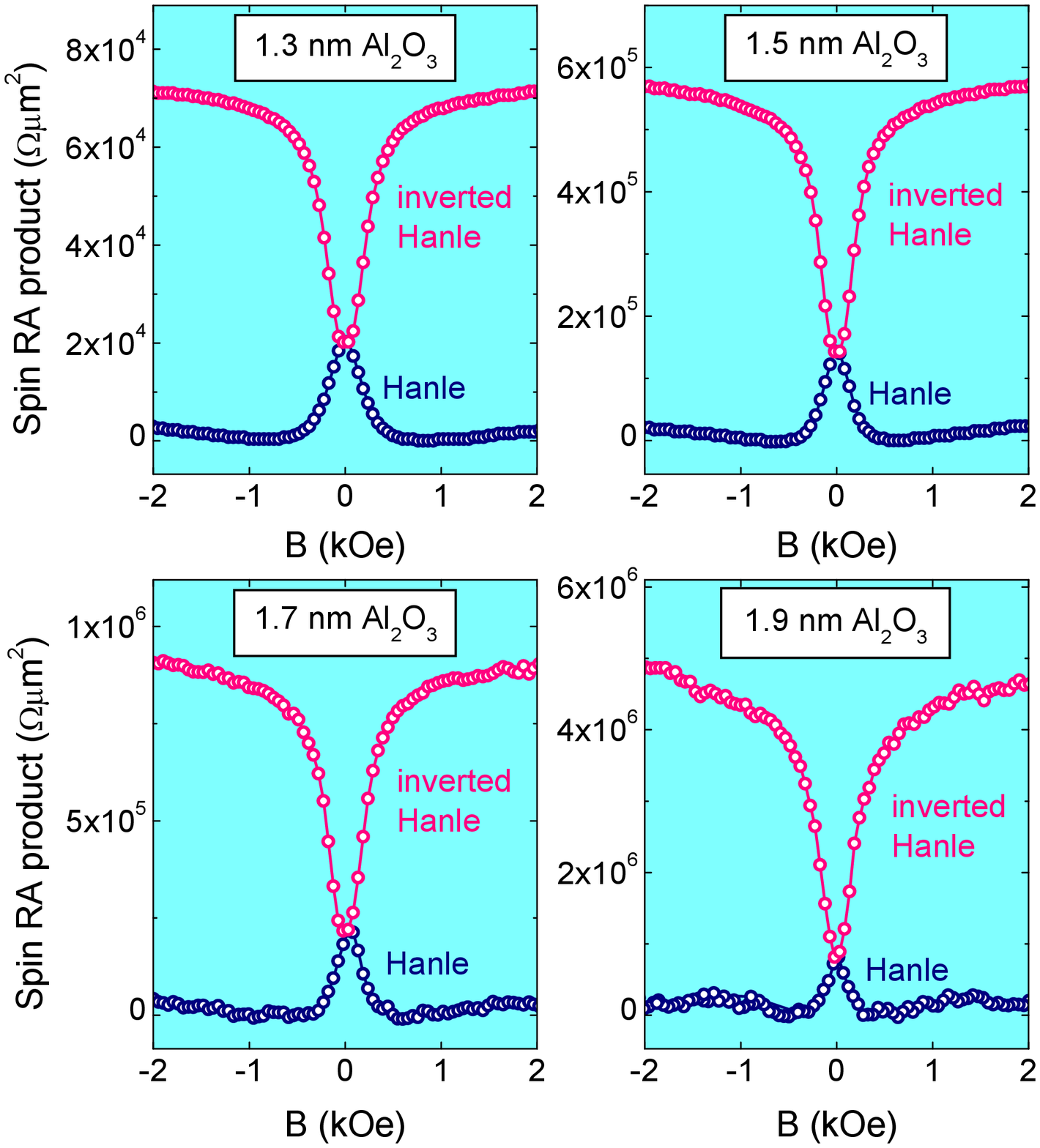}\hspace*{0mm}\includegraphics*[width=70.7mm]{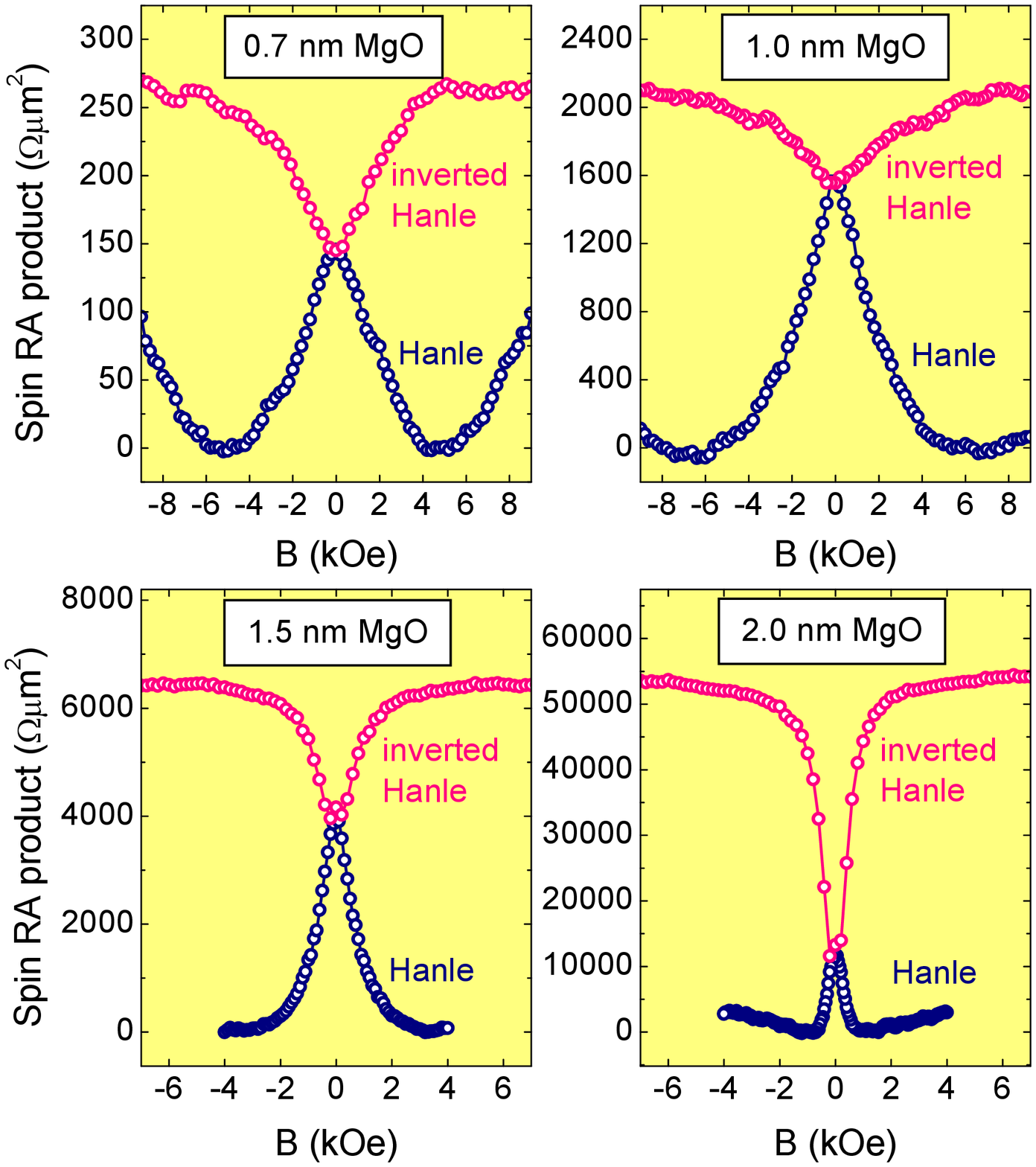}
\caption{Hanle detection of spin accumulation in
semiconductor/oxide/ferromagnet tunnel devices. Shown are
representative Hanle and inverted Hanle curves for magnetic field
$B$ applied, respectively, perpendicular or parallel to the
magnetization of the ferromagnet. Data is shown for p-type
Si/Al$_2$O$_3$/Ni$_{80}$Fe$_{20}$ and p-type Si/MgO/Fe devices
with different thickness of the tunnel barrier, all at 300 K. The
vertical axis gives the spin-RA product, defined as $\Delta
V_{Hanle}/J$.} \label{fig2}
\end{figure}

\indent The tunnel resistance exhibits the expected exponential variation with thickness of the
tunnel oxide (Fig. 2a). From the slope we extract an effective tunnel barrier height $\Phi_{eff}$ of 0.8 eV.
Taking into account the effective electron mass in Al$_2$O$_3$ (about 0.2 - 0.3 times the free
electron mass), this translates into a real barrier height of $\Phi=$ 3.2 $\pm$ 0.8 eV. This is a
reasonable value \cite{robertson} for Al$_2$O$_3$ on p-type Si, showing that direct
tunneling from the ferromagnet into the Si is the dominant transport process (for multi-step
tunneling via localized states within the oxide \cite{beasley}, the extracted barrier height
would be 4 times larger, which is unrealistic). More importantly, the data implies that
the contact resistance is dominated by the Al$_2$O$_3$, and that the depletion region
associated with the Schottky barrier in the Si contributes little to the resistance, as
expected for heavily doped Si.

\begin{figure}[htb]
\hspace*{0mm}\includegraphics*[width=40mm]{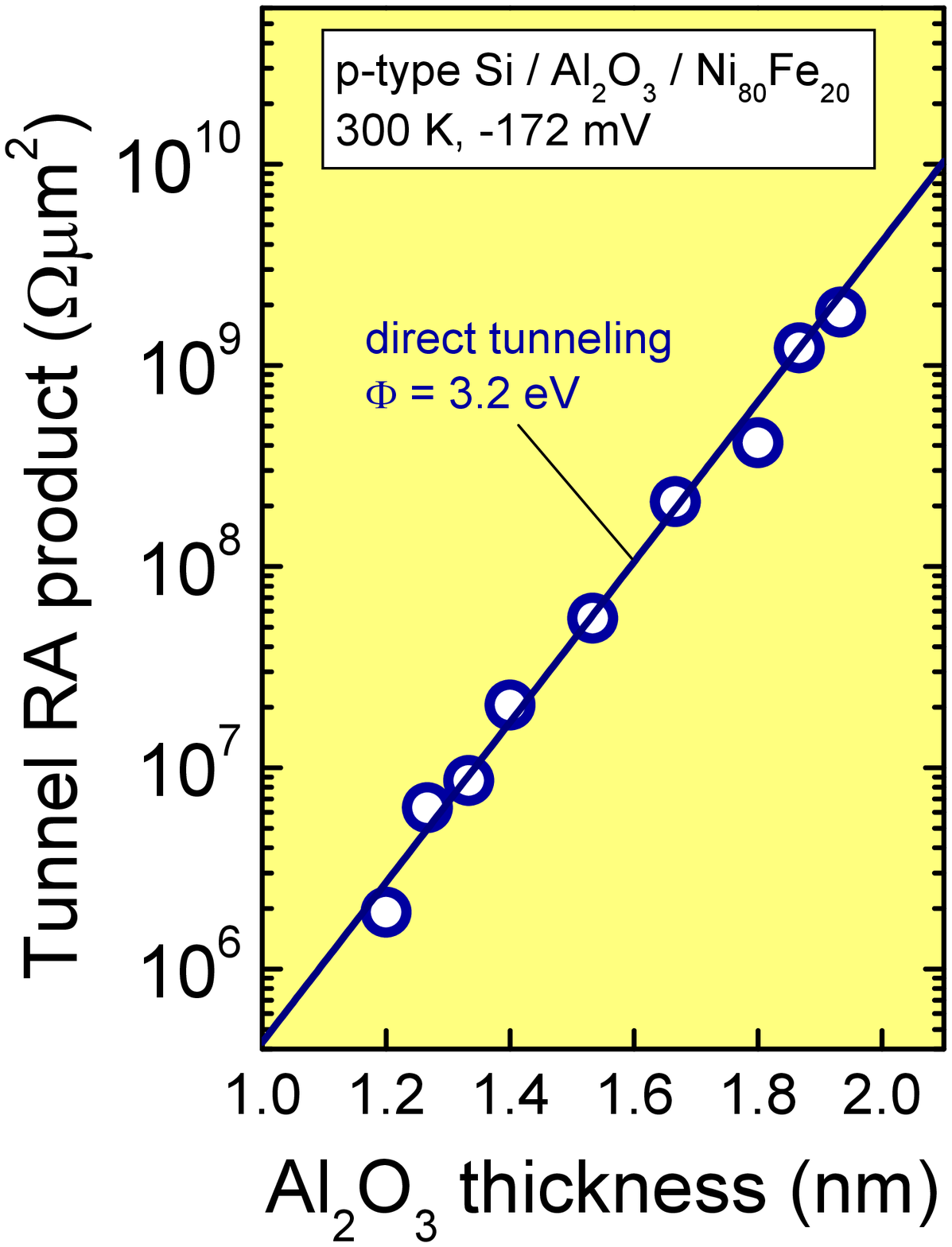}
\hspace*{0mm}\includegraphics*[width=68.5mm]{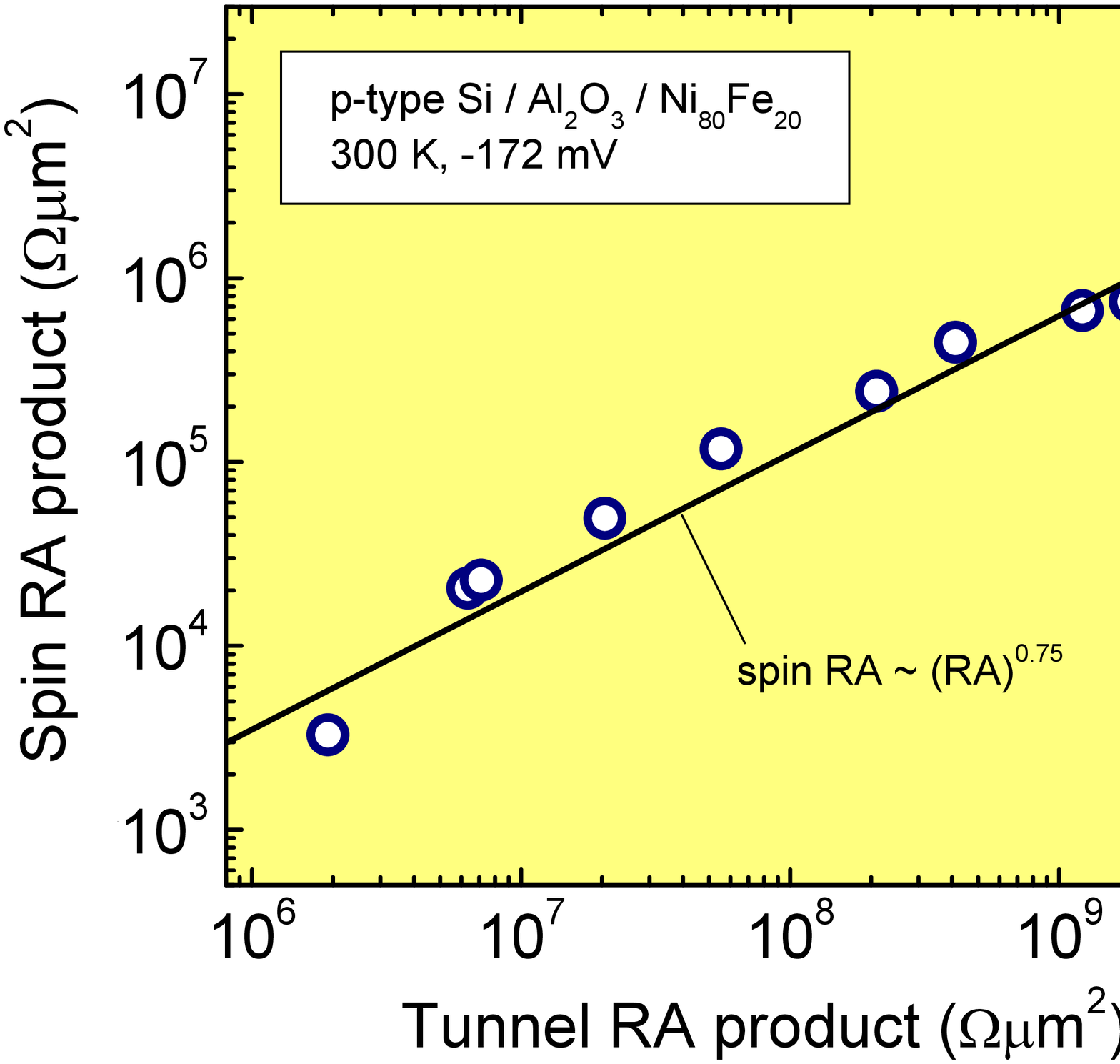}
\hspace*{0mm}\includegraphics*[width=68mm]{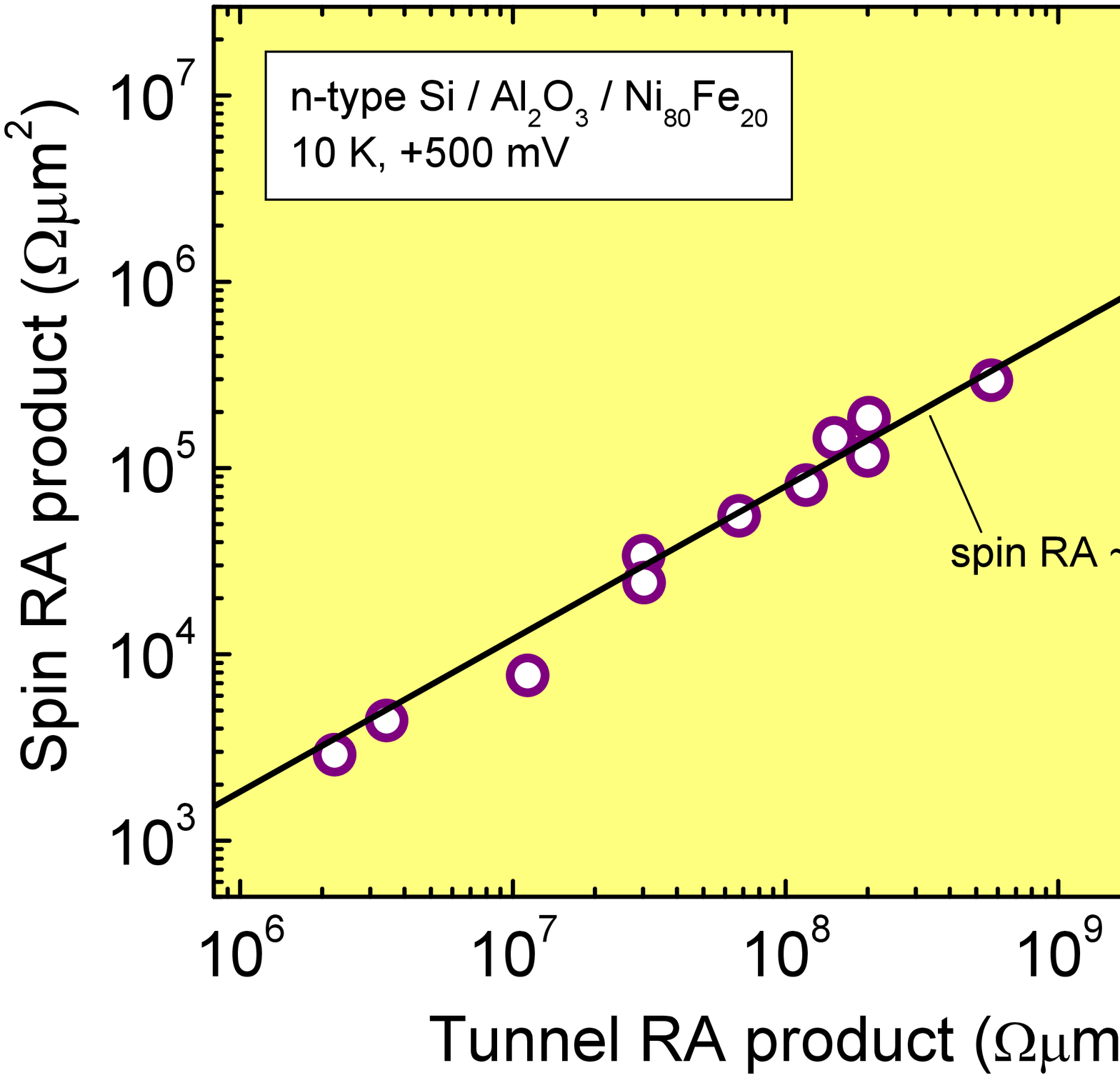}
\caption{Scaling of Hanle spin signals in Si tunnel devices with
amorphous Al$_2$O$_3$ barrier. (a), tunnel resistance-area (RA) product
versus Al$_2$O$_3$ thickness for p-type Si/Al$_2$O$_3$/Ni$_{80}$Fe$_{20}$
devices at 300 K. The extracted tunnel barrier height
is 3.2 eV. (b), the corresponding spin RA product versus tunnel RA
product. The solid line corresponds to a power law with exponent 0.75. (c),
data for similar devices but with n-type Si at 10 K. The solid line
corresponds to a power law with exponent 0.82. The spin
RA value is derived from the Hanle signal only, instead of the sum
of the Hanle and inverted Hanle signals. See appendix A
for additional data with different oxidation time (p-type) and Cs treated surfaces
(n-type).}
\label{fig3}
\end{figure}

\begin{figure}[htb]
\hspace*{0mm}\includegraphics*[width=54mm]{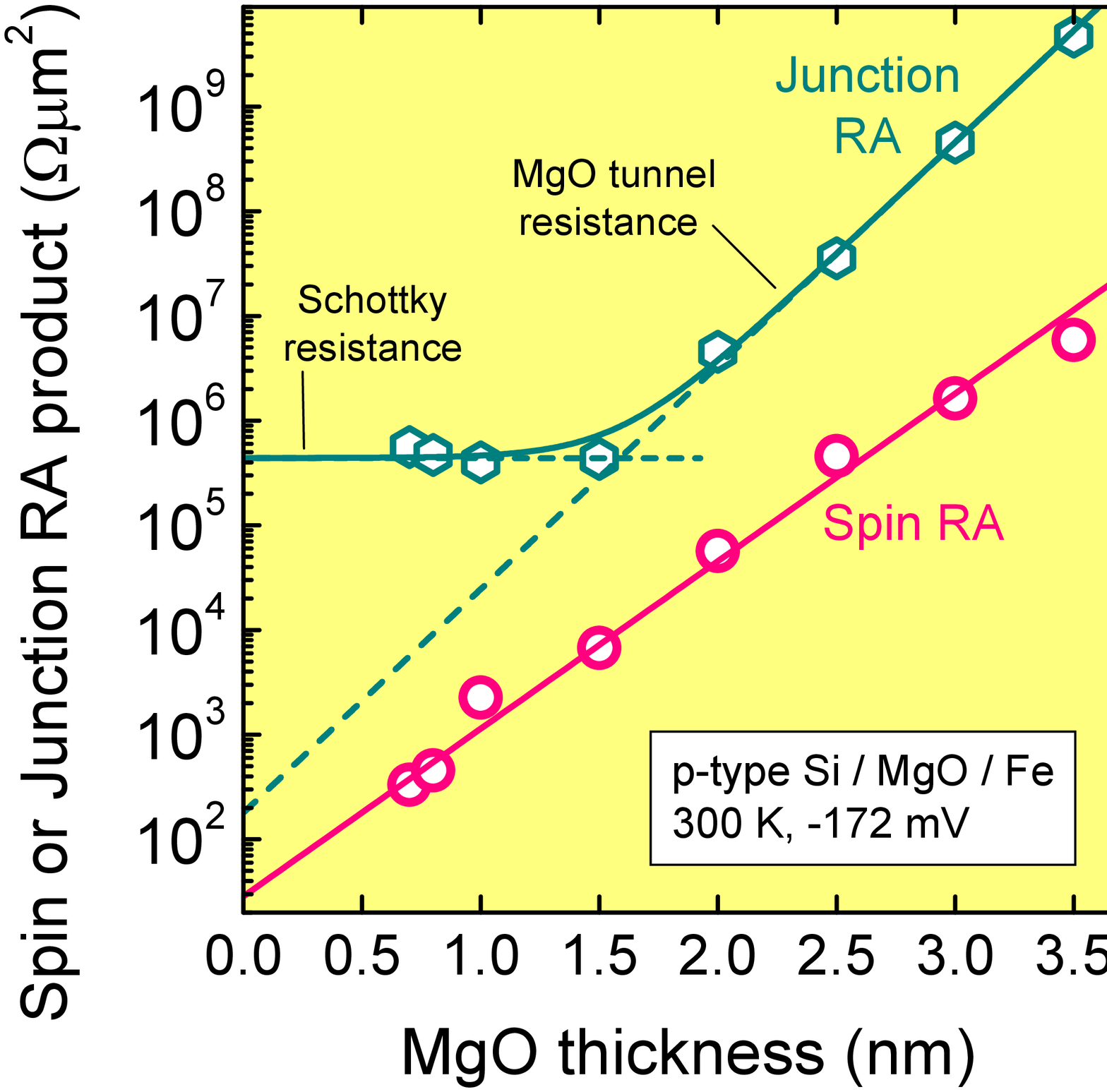}
\hspace*{0mm}\includegraphics*[width=68mm]{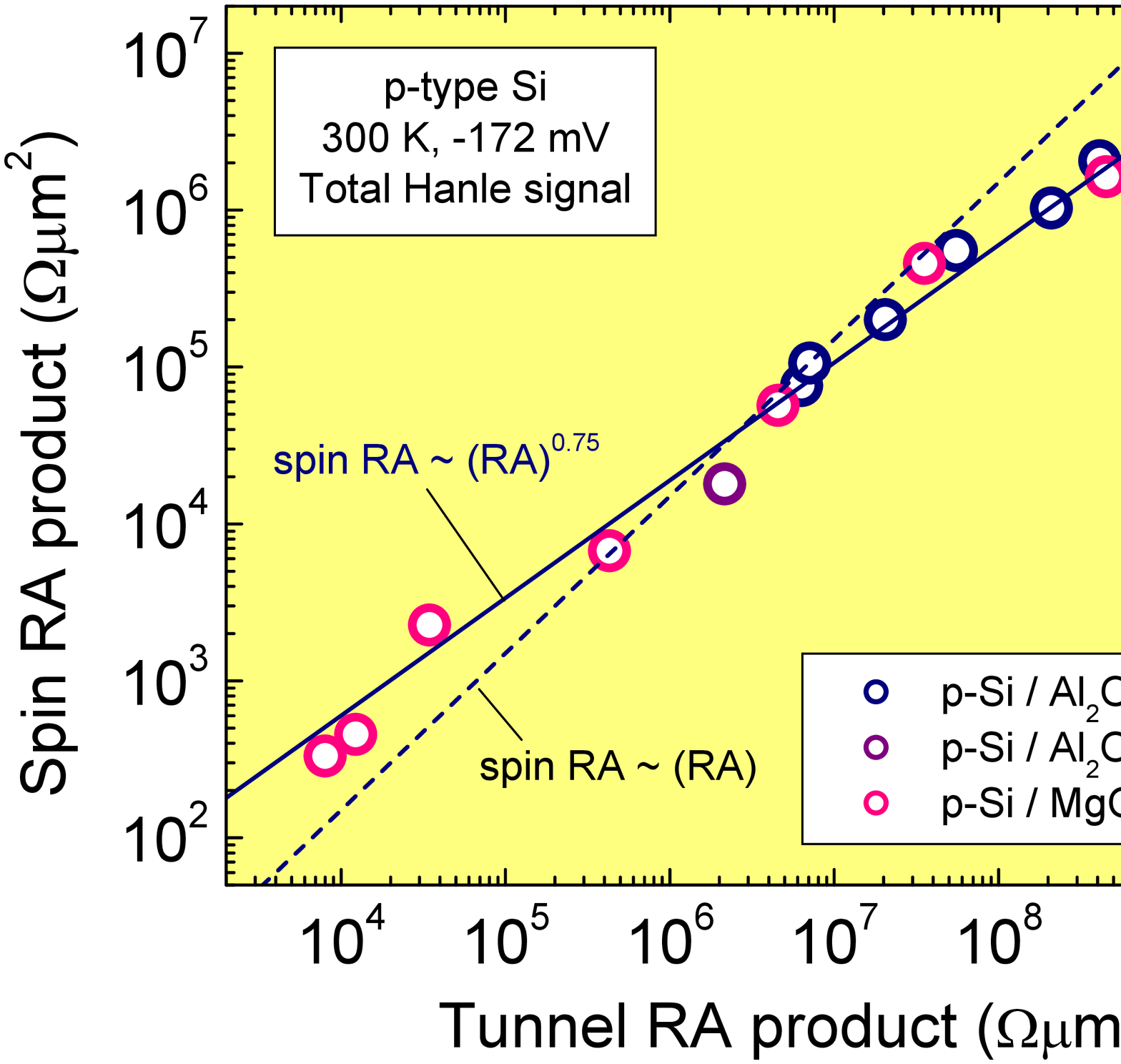}
\hspace*{0mm}\includegraphics*[width=53mm]{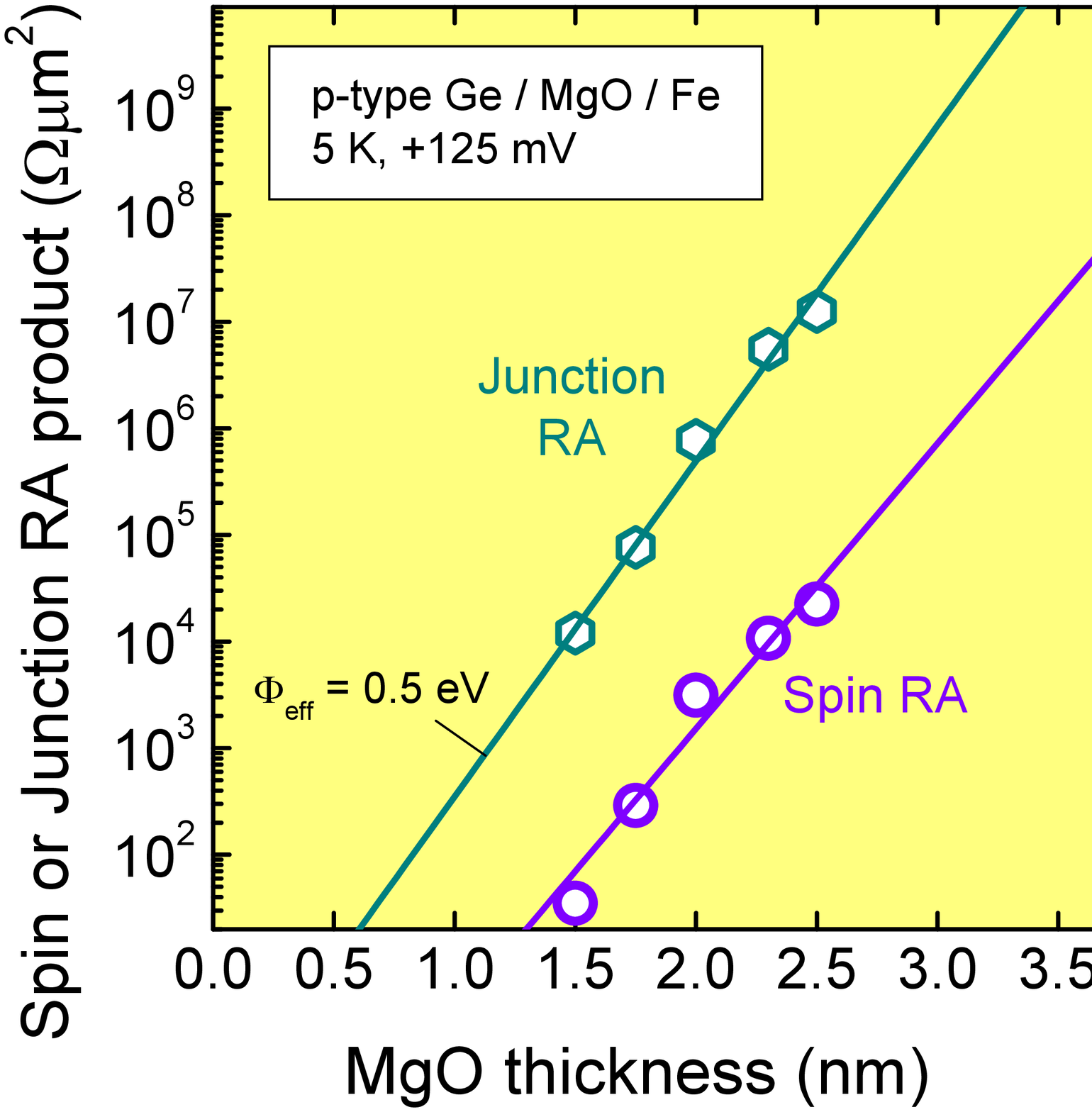}
\caption{Scaling of Hanle spin signals for devices
with MgO and p-type Si or Ge. (a), spin RA product and tunnel RA product
versus MgO thickness for p-type Si/MgO/Fe devices at 300 K. (b),
corresponding spin RA product versus tunnel RA product for the
same devices (pink symbols), together with data
for p-type Si/Al$_2$O$_3$/Ni$_{80}$Fe$_{20}$ (blue symbols). For the 3
devices with the thinnest MgO barrier, the tunnel RA product is
determined by extrapolation from the high thickness regime (dashed
green line in {\bf a}) to remove the Schottky resistance. The
solid line corresponds to a power law with exponent 0.75. (c),
data for p-type Ge/MgO/Fe devices at 5 K. The spin
RA product is the sum of the Hanle and inverted Hanle signal.}
\label{fig4}
\end{figure}

\indent The spin RA product also displays an exponential variation with thickness of the
tunnel oxide, and a power law is revealed when the spin RA product is plotted against tunnel
resistance (Figs. 2b and 2c). The associated scaling exponent is about 0.75 and 0.82,
respectively, for Si/Al$_2$O$_3$/Ni$_{80}$Fe$_{20}$ devices with p-type and n-type Si. For
devices with crystalline MgO/Fe contacts, a similar exponential variation of spin RA product
with MgO thickness is obtained (Fig. 3a). The contact resistance is dominated by tunneling
through the MgO at larger thickness, but for small MgO thickness a transition occurs to
the regime where the contact resistance is limited by the Schottky barrier and becomes constant.
Interestingly, the spin RA product displays no transition. It scales with the MgO thickness even
in the low thickness regime, suggesting that the spin signal is determined by the tunneling across
the MgO. The spin RA products for p-type Si with MgO/Fe and Al$_2$O$_3$/Ni$_{80}$Fe$_{20}$ contacts
display similar scaling as a function of resistance of the tunnel oxide (Fig. 3b), with an exponent
(0.75) smaller than 1. For devices on heavily doped p-type Ge with crystalline MgO/Fe contacts \cite{ibage,ibagert},
we also find that $R_{tun}$ and the spin RA product vary exponentially with MgO thickness (Fig. 3c),
although the data set is too limited to extract an accurate value for the scaling exponent. Note that
a similar scaling was recently reported by Uemura {\it et al.} for n-type Si/MgO/Co$_{50}$Fe$_{50}$ devices \cite{uemura} ,
although a direct comparison cannot be made because their data was taken with the same bias current
for each oxide thickness, and hence with a different tunnel voltage. This, in turn, changes
the tunnel spin polarization, which is known to vary with the energy of the tunnel electrons \cite{valenzuela,park}.
This additional source of variation of the spin signal with tunnel oxide thickness is not present in our data,
which was obtained using the same bias voltage for each oxide thickness. Our collection of data
leads to the striking and unexpected conclusion that $\Delta V_{Hanle}/J$ is not a constant but
scales with $R_{tun}$, and up to values larger than 10$^{9}$ $\Omega\mu m^2$. This behavior is generic,
as it is observed for devices with different semiconductors, tunnel oxides, and ferromagnetic electrodes.
Below we explain that this behavior is incompatible with any of the known theories for the injection,
accumulation and diffusion of spins in ferromagnetic tunnel devices.

\section{Comparison with existing theory}
\subsection{Direct tunneling}
\indent In the standard theory, the spin accumulation gives rise to a Hanle spin
signal $\Delta V_{Hanle}=[P_{fm}^2]\,r_s\,J$, where $P_{fm}$ is the tunnel spin
polarization associated with the oxide/ferromagnet interface, and $r_s$ is the
spin resistance of the semiconductor \cite{jansensstreview,fabianacta} that
describes the relation between spin current and spin accumulation in non-magnetic
materials. Thus, $\Delta V_{Hanle}/J$ is constant and independent of the resistance
$R_{tun}$ of the tunnel contact. This applies when $R_{tun}$ is larger than
$r_s$. If $R_{tun}<r_s$, back flow of the spins into the
ferromagnet limits the spin signal \cite{jansensstreview,fertPRB,fabianacta}, which is then proportional to
the tunnel resistance: $\Delta V_{Hanle}/J=[P_{fm}^2/(1-P_{fm}^2)]\,R_{tun}$.
Although this produces a scaling with tunnel resistance, the experimentally
observed scaling extends to tunnel RA values beyond 10$^{9}$
$\Omega\mu m^2$, and a value of $r_s$ larger than this would be
required for back flow to be active. This is unreasonable, since
$r_s$ is typically around 10-100 $\Omega\mu m^2$ for the semiconductors
used \cite{jansensstreview}. The standard description thus predicts that
$\Delta V_{Hanle}/J$ is independent of $R_{tun}$ (and up to 7 orders
of magnitude smaller than observed) and cannot describe the data.

\subsection{Two-step tunneling}
\indent Two-step tunneling via localized states near the oxide/semiconductor interface
can produce an enhanced spin signal due to spin accumulation in those states \cite{tran},
provided that certain conditions are satisfied \cite{jansensstreview}. To obtain a spin resistance
of 10$^{9}$ $\Omega\mu m^2$, localized states with a spin lifetime of at least 10 $\mu s$
are needed for a reasonable density of interface states ($> 10^{12}$ states$/eVcm^2$).
While this cannot be excluded a priori, there exists ample experimental evidence
that shows that this mechanism is not the origin of the large spin signals
observed, as recently reviewed \cite{jansennmatreview,jansensstreview}.
The scaling data presented here provides additional and conclusive proof
that two-step tunneling via interface states cannot be responsible,
as it leads to a fundamental inconsistency. The scaling of the
contact resistance with oxide thickness (Fig. 2a) implies that the
resistance is dominated by the oxide tunnel barrier, and that any
resistance $r_b$ of the Schottky barrier in the semiconductor is
much smaller. For two-step tunneling, the effective spin
resistance $r_s^{eff}$ of the interface states cannot be larger
than the resistance $r_b$ that couples the states to the bulk
semiconductor \cite{jansensstreview,tran}. Taken together this would
mean $r_s^{eff}<r_b<R_{tun}$. However, in order to obtain a scaling of
the spin RA with tunnel resistance (due to back flow from the
interface states into the ferromagnet), one needs the opposite,
namely, $R_{tun}<r_s^{eff}$. These requirements cannot be
satisfied simultaneously, whatever the parameters chosen. Thus,
Tran's model \cite{tran} for spin accumulation in interface states
is inconsistent with the simultaneous exponential scaling of
contact resistance and spin signal with tunnel barrier thickness.

\subsection{Two-step and direct tunneling in parallel}
\begin{figure}[htb]
\hspace*{0mm}\includegraphics*[width=52mm]{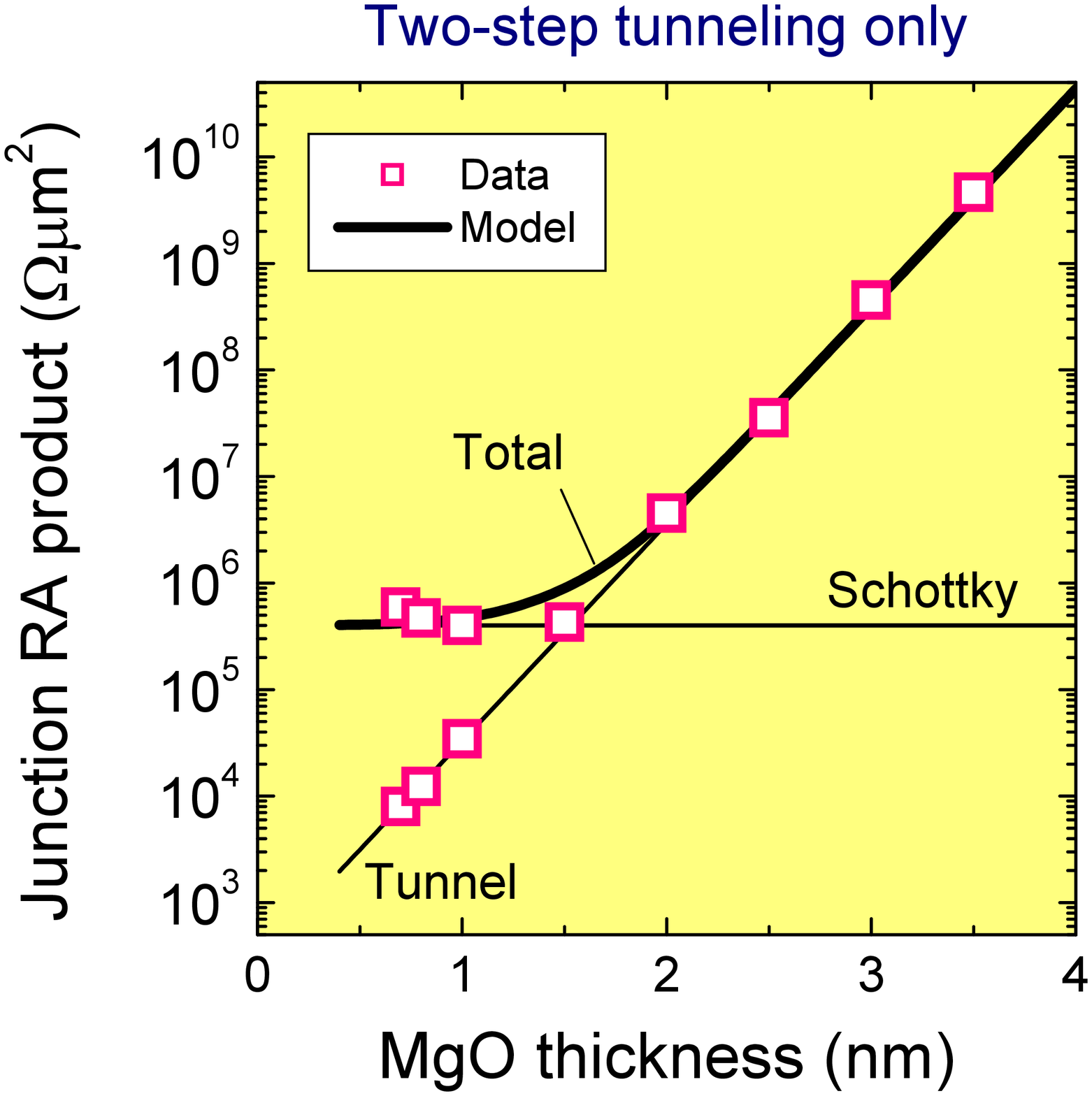}\hspace*{10mm}\includegraphics*[width=52mm]{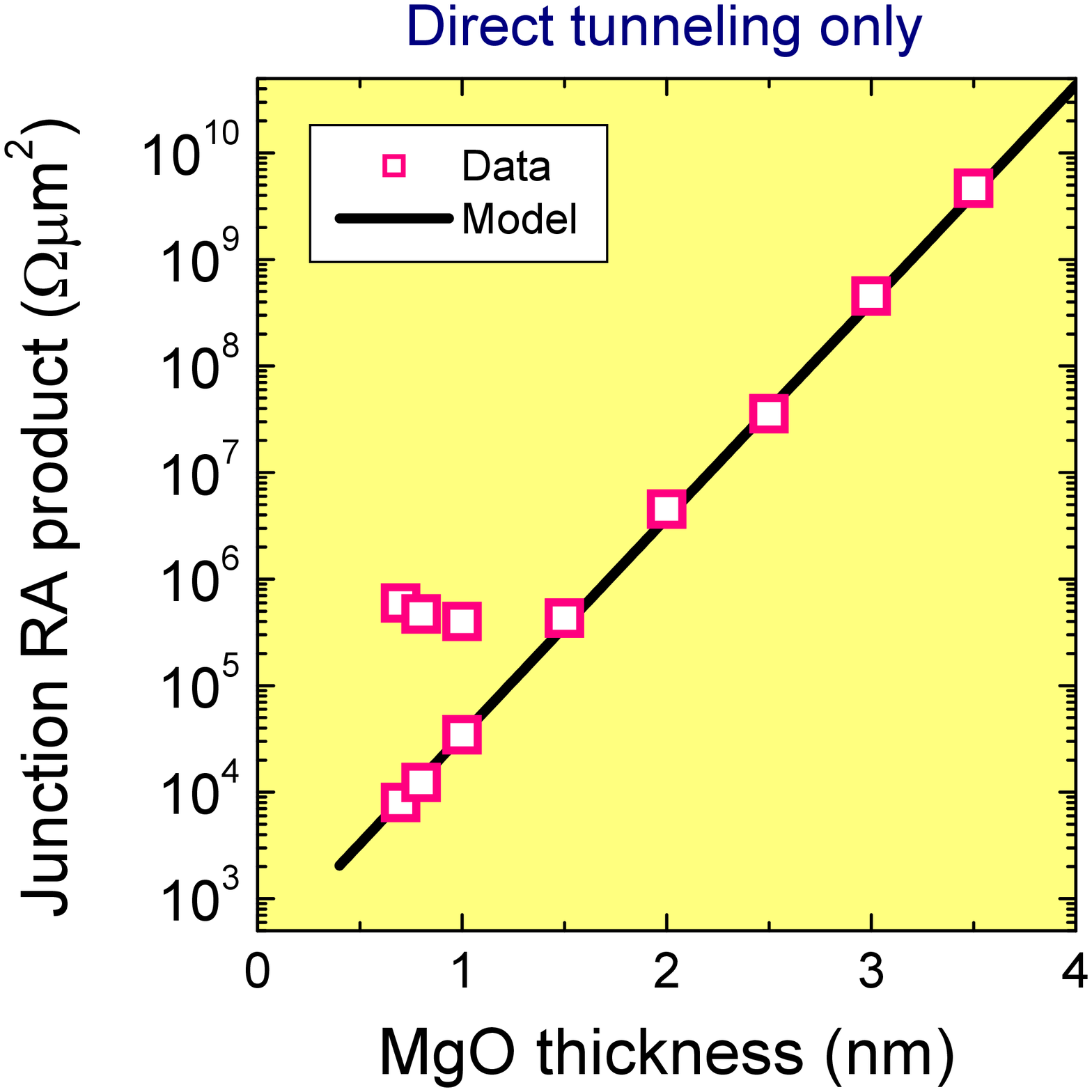}\hspace*{10.1mm}\includegraphics*[width=52mm]{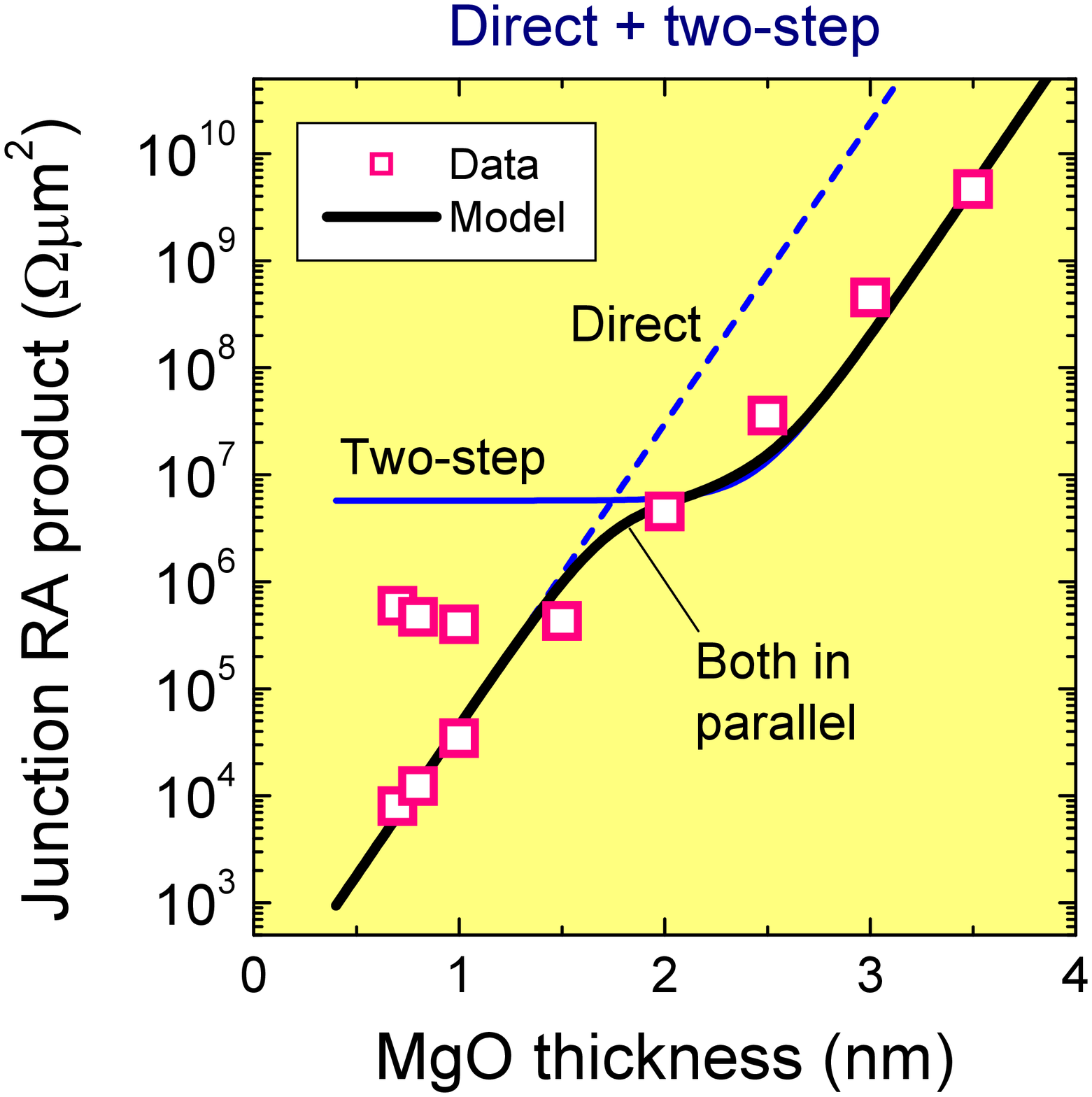}\\
\vspace*{-1mm}\hspace*{0mm}\includegraphics*[width=52mm]{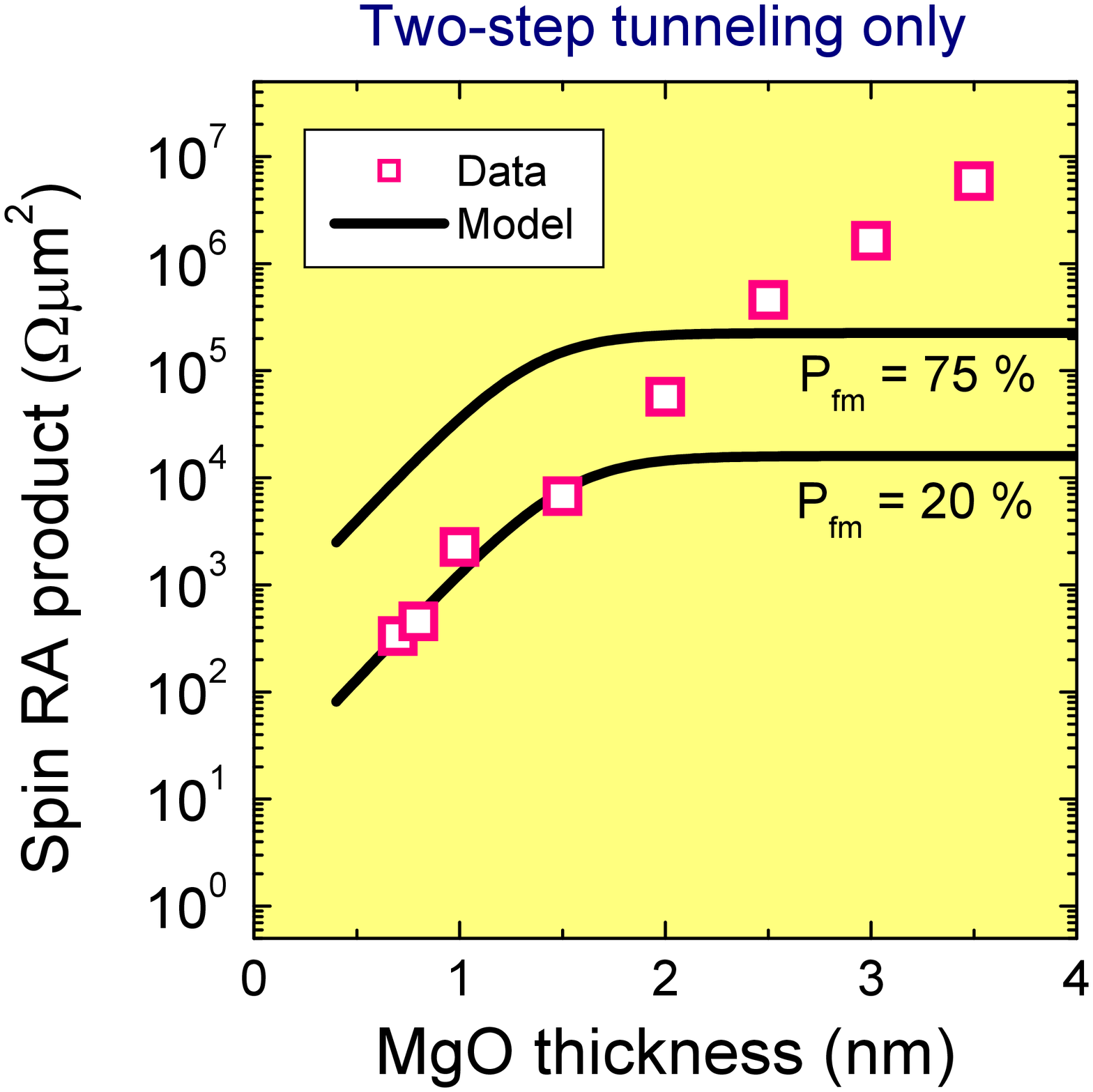}\hspace*{10.3mm}\includegraphics*[width=51.6mm]{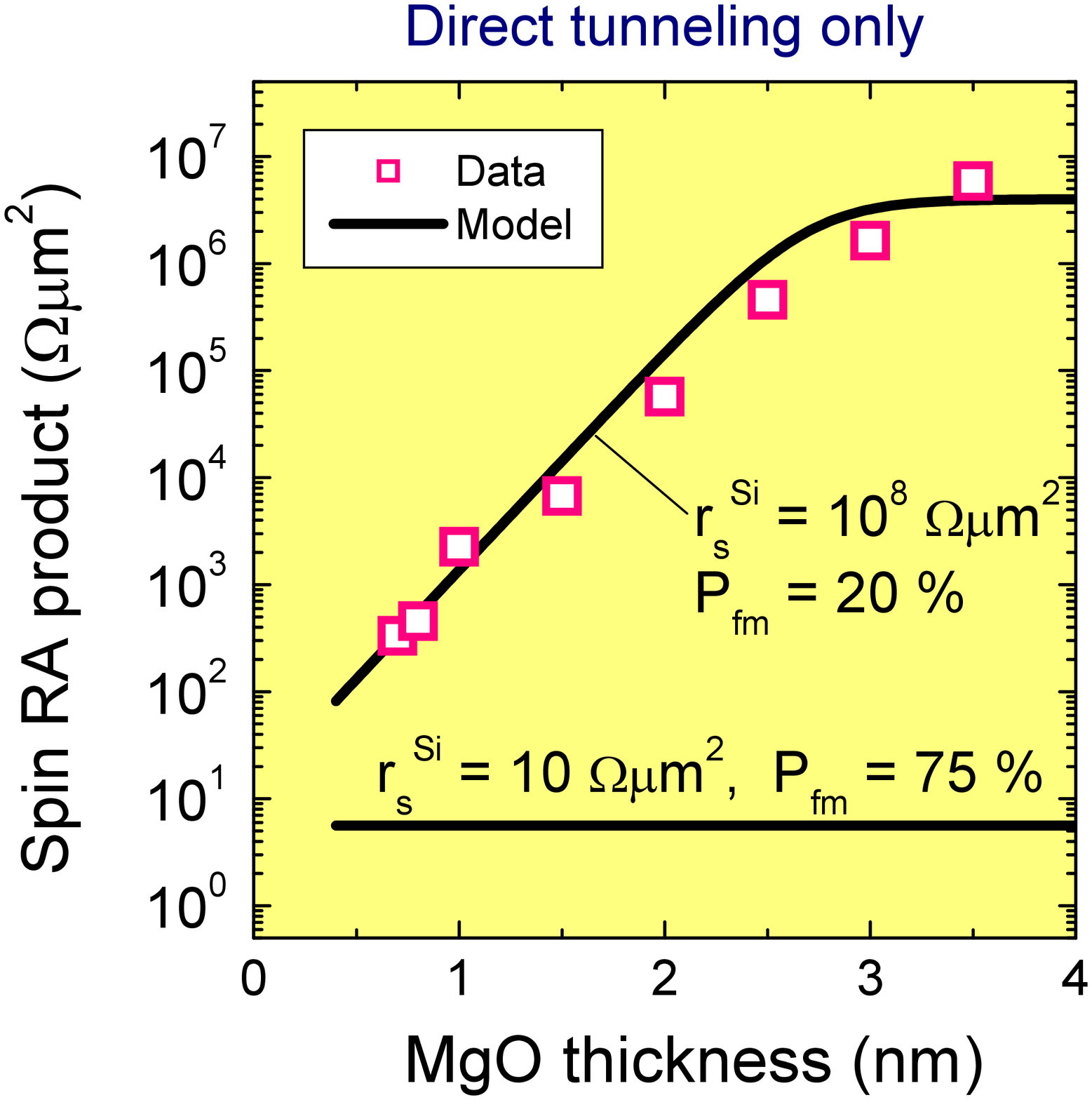}\hspace*{10.1mm}\includegraphics*[width=52.1mm]{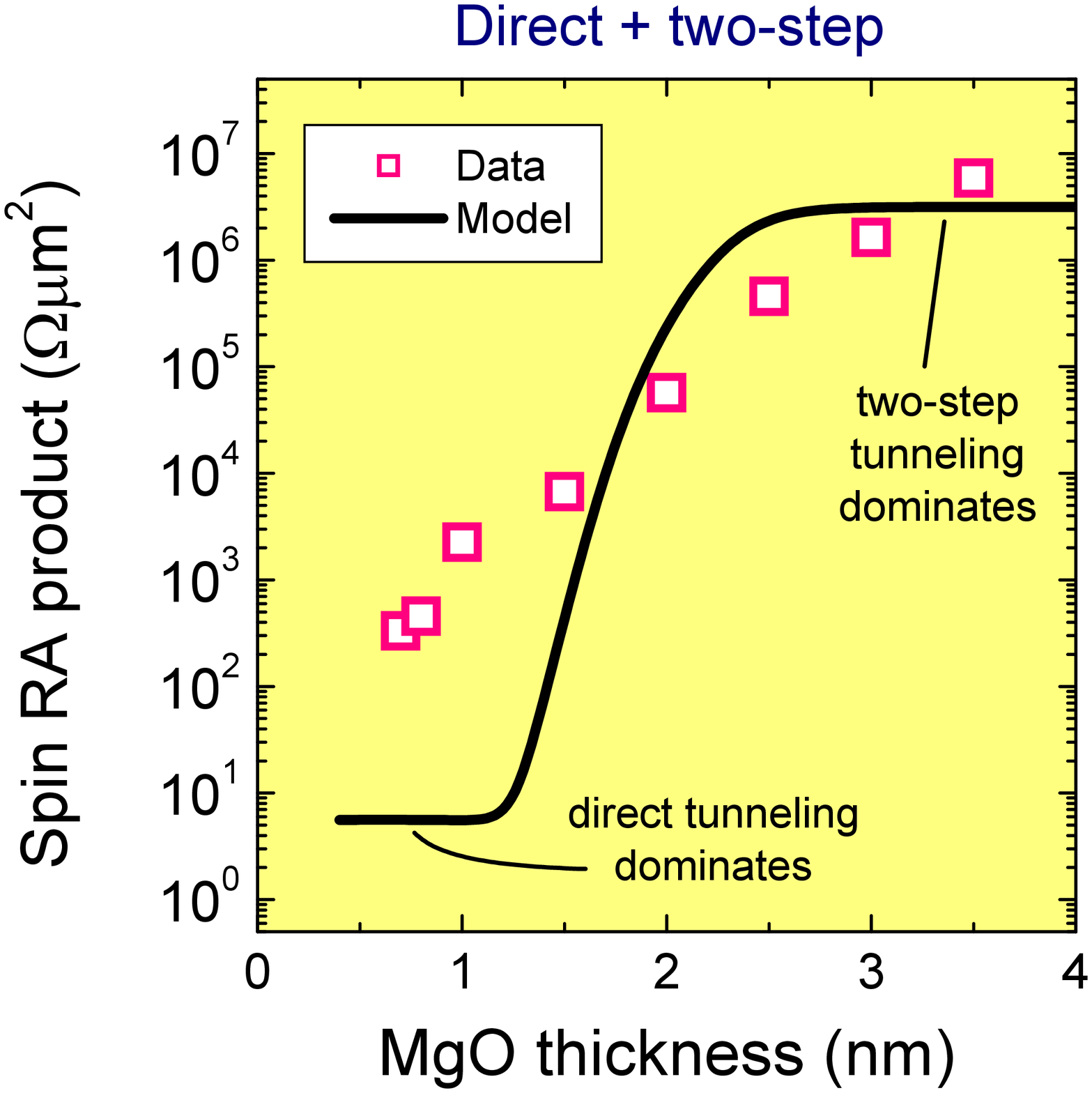}
\caption{Attempts to fit the data by direct and two-step tunneling in parallel. The experimental data (pink symbols) for the junction resistance and spin RA product of p-type Si/MgO/Fe devices is compared with the model (solid lines) for three cases: (i) two-step tunneling only, (ii) direct tunneling only, and (iii) two-step tunneling and direct tunneling in parallel. In the top panels, two data points are given for the junction RA product of the three junctions with smallest MgO thickness; the larger value corresponds to the measured junction RA product, whereas the smaller value is obtained when the resistance of the Schottky barrier is subtracted.}
\label{figS2}
\end{figure}

\indent The model introduced by Tran et al. \cite{tran} for two-step tunneling via localized states assumes that {\em all} the current goes via the localized states and, as just noted, this cannot describe the experimental data. However, Tran's model has recently been extended \cite{jansentwostep} by including charge and spin transport by direct tunneling, in parallel with two-step tunneling. It is therefore important to examine whether this extended transport model can describe the experimental data. Depending on the details of the system, the spin accumulation created in localized states due to two-step tunneling can be much larger than that induced in the semiconductor bands by direct tunneling. If as a function of some parameter (e.g. the tunnel barrier thickness) the relative contribution of two-step and direct tunneling is changed, then a transition from a small signal (direct tunneling dominant) to a large signal (two-step tunneling dominant) or vice versa, can be produced. In this paragraph we examine whether this can explain the observed scaling of the spin RA product with tunnel barrier thickness, and show that this is not possible.\\
\indent We start by attempting to fit the experimental data for the p-type Si/MgO/Fe devices by setting the direct tunnel current to zero. That is, we consider transport by two-step tunneling, where the first step is by tunneling across the MgO from ferromagnet into interface states, and the second step is by tunneling through the Schottky barrier from interface states into the bulk semiconductor. The result
is shown in the left two panels of Fig. \ref{figS2}, for which the spin resistance of the localized states was set to infinity, so that the spin signal is not limited by spin relaxation in the localized states. A good fit (thick solid lines) is obtained for the junction resistance. For small MgO thickness, the junction resistance is limited by the resistance of the Schottky barrier, whereas at large MgO thickness it is limited by tunneling across the MgO. According to the model, this should be accompanied by a transition in the behavior of the spin RA product, which first increases with MgO thickness, but becomes constant as soon as the junction resistance is determined by the MgO. This corroborates the statement made in the previous paragraph that the junction resistance and spin resistance cannot simultaneously exhibit a scaling with MgO thickness if transport is by two-step tunneling via localized interface states. Note that in the regime of small MgO thickness, the magnitude of the spin signal is determined by the tunnel spin polarization $P_{fm}$ associated with the Fe/MgO interface, and a value of 20\% is needed to obtain a match with the data in this regime. With a value of 75\%, which is more reasonable \cite{yuasa}, the data cannot be described, not even in the regime of small MgO thickness.

\indent The two middle panels show the result if transport is purely by direct tunneling, setting the two-step tunnel current to zero. In principle the data can be described, however, the required spin resistance $r_s^{Si}$ of the silicon is of the order of 10$^8$ $\Omega\mu$m$^2$. This is unreasonable considering that is expected to be in the range of 10 - 100 $\Omega\mu$m$^2$ at best, for which one would obtain a spin RA product that is independent of the tunnel oxide thickness and orders of magnitude smaller than experimentally observed.\\
\indent Next, we attempt to describe the data by direct and two-step tunneling in parallel, using the equations given in Ref. \onlinecite{jansentwostep}. As already eluded to above, in order to obtain an increase of the spin RA product as a function of MgO thickness, one needs to have a transition from transport dominated by direct tunneling to transport dominated by two-step tunneling via interface states. Such a situation is depicted in the two right panels of Fig. \ref{figS2}. At large MgO thickness, transport is determined by tunneling across the MgO, and we have chosen the parameters such that in this regime the resistance associated with two-step tunneling is smaller than that for direct tunneling. At small MgO thickness, the two-step tunnel current is limited by the resistance of the Schottky barrier. As a result, the transport at small MgO thickness is dominated by direct tunneling. This change in transport process can reasonably well describe the observed scaling of the junction resistance, but not the scaling of the spin signal. A transition from a small spin RA product, governed by direct tunneling, to an enhanced spin RA product due to two-step tunneling is indeed created, but the model does not reproduce the experimental data. It does not reproduce the observed exponential increase of the spin RA with MgO thickness, and deviates from the data in almost the entire range. We conclude that a transition in transport from direct to two-step tunneling does not describe the experimental data.

\subsection{Inhomogeneous tunnel current density}
\indent It has previously been pointed out that an enlarged spin signal can be produced in
three-terminal devices if the tunnel current density is not homogeneous across the
contact area \cite{dash}. In that case the local current density, and thereby the spin
accumulation, can be significantly larger than what is expected from the applied current
and the lateral dimensions of the tunnel contact. In previous work \cite{dash}
the spin signal was larger than expected by 2-3 orders of magnitude and in principle
this could be due to lateral inhomogeneity of the tunnel current. However, the new data
presented here exhibit a scaling with tunnel barrier resistance that is not readily
understandable with an explanation in terms of current inhomogeneity. Moreover, for
devices with the thickest tunnel barrier, the observed spin signals are larger than expected
by up to 6 orders magnitude, and this cannot be explained by inhomogeneous tunnel current.
It would require that all the tunnel current goes via an area that is 10$^6$ times smaller
than the geometric contact area of 100 $\times$ 200 $\mu$m$^2$. This translates into an
effective tunnel area of only 100 $\times$ 200 nm$^2$ or so, which is unreasonable. We
conclude that inhomogeneity of the tunnel current is not responsible for the experimental
observations.

\section{Control devices}
\subsection{Devices with metal instead of semiconductor}
\begin{figure}[htb]
\hspace*{0mm}\includegraphics*[width=54.2mm]{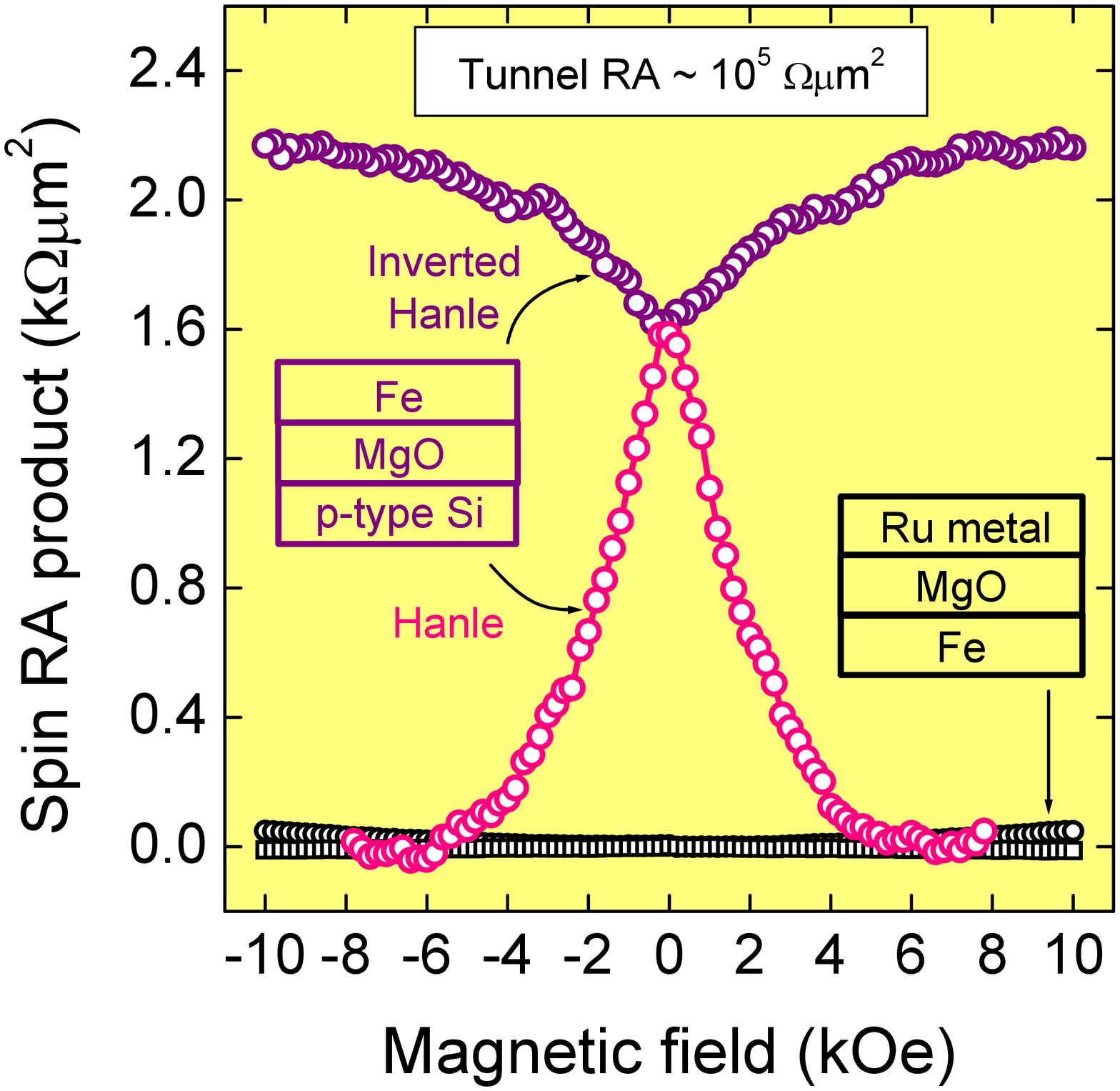}
\hspace*{0mm}\includegraphics*[width=55mm]{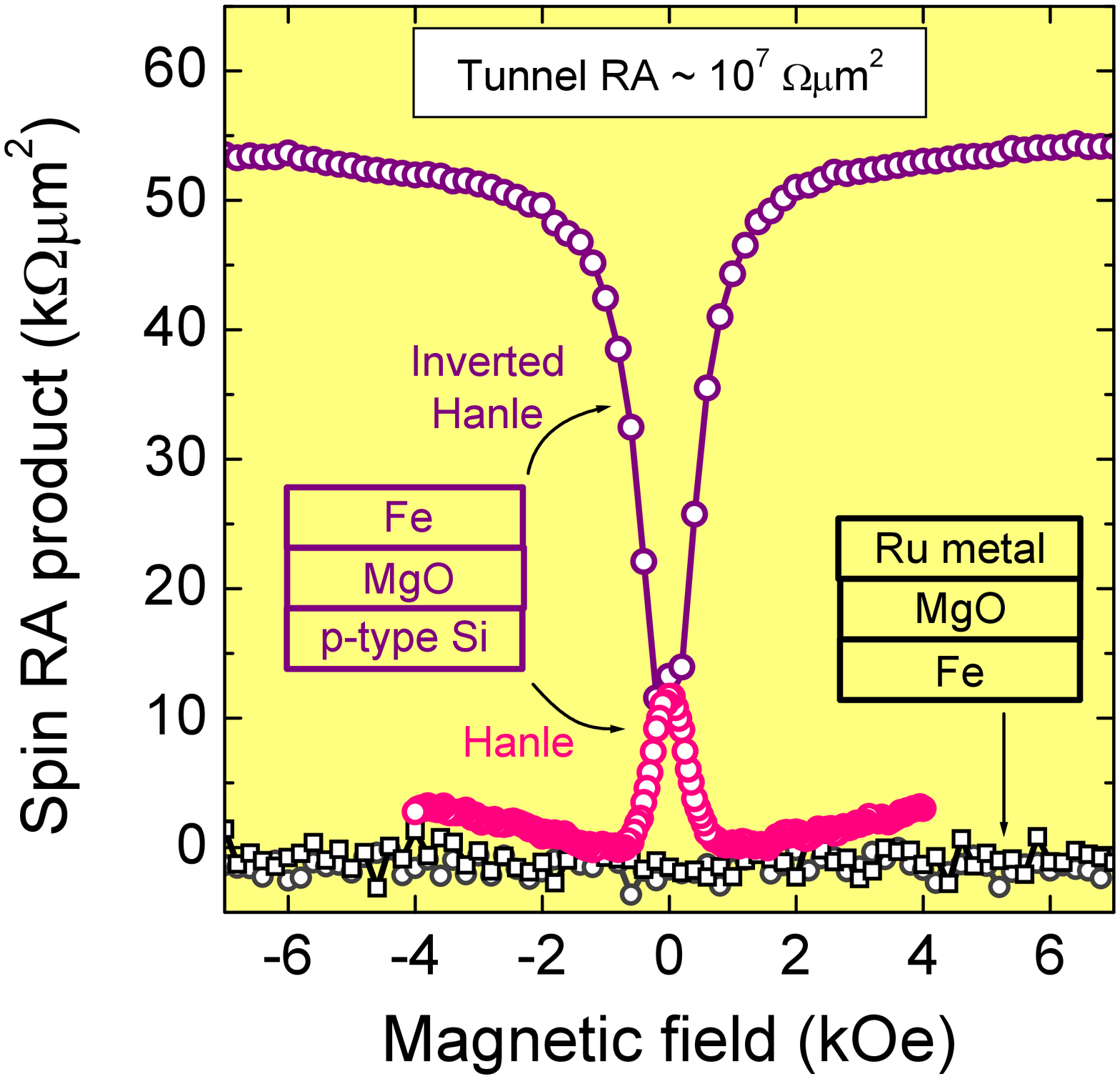}
\caption{Absence of spin accumulation in control tunnel
devices with the Si replaced by non-magnetic Ru metal. Shown are
the Hanle and inverted Hanle signal for p-type Si/MgO/Fe devices,
together with similar data on control devices with Fe/MgO/Ru
structure (black circles and squares). Measurements were done at
300 K, and devices with similar tunnel RA product are compared
(10$^{5}$ $\Omega\mu m^2$ and 10$^{7}$ $\Omega\mu m^2$ for the
left and right panel, respectively). The absence of any spin
signal in the metallic control devices proves the absence of spin
accumulation in localized states within the MgO tunnel barrier.}
\label{figS1}
\end{figure}

\indent The argument used in the previous section to rule out two-step tunneling via localized interface states was based on the assumption that the states are located at the oxide/semiconductor interface, and decoupled from the semiconductor bulk bands by a Schottky barrier with resistance $r_b$. In principle, it is possible that the relevant localized states are present {\em within} the oxide tunnel barrier, and that a large spin accumulation is induced in those states by two-step tunneling. In this case the value of $r_b$ that couples the localized states to the semiconductor bands is no longer determined by the Schottky barrier, but by the resistance of part of the tunnel oxide. It is known that two-step tunneling is more efficient for states near the center of the tunnel barrier \cite{beasley}. Hence, the associated value of $r_b$ is determined by half of the tunnel oxide and would systematically increase with the thickness of the tunnel oxide. Depending on the parameters of the system, this could produce a spin accumulation that increases with tunnel barrier thickness, and thereby a scaling of the spin RA product with tunnel resistance.\\
\indent In order to exclude this possibility, we fabricated control devices in which the semiconductor is replaced by a non-magnetic metal (Ru) electrode. If the large spin RA product originates from spin accumulation in states within the tunnel oxide, the spin accumulation does not depend on the spin resistance of the non-magnetic electrode, and a similarly large spin accumulation should be observed with a Ru metal electrode. However, in control devices with the structure Fe/MgO/Ru, no spin signal could be observed, neither Hanle nor inverted Hanle (see Fig. \ref{figS1}). Therefore, we conclude that the spin signal does not originate from spin accumulation in localized states in the tunnel oxide. This control experiment also rules out the recent proposal \cite{uemura} of spin accumulation in states localized in the tunnel barrier close to the oxide/ferromagnet interface, in which a large spin accumulation is not expected to exist anyway because the strong coupling with the ferromagnet would easily deplete the spin accumulation.\\

\subsection{Devices with zero tunnel spin polarization}
\begin{figure}[htb]
\hspace*{0mm}\includegraphics*[width=62mm]{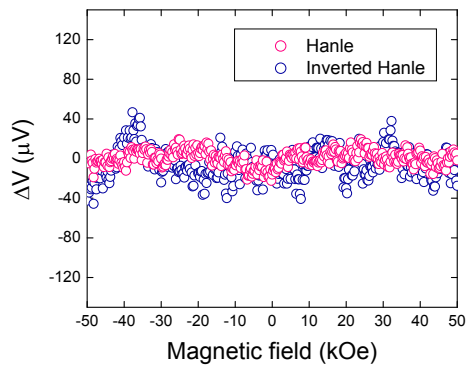}
\caption{Absence of spin signals in control tunnel
devices with zero tunnel spin polarization. Shown are
the Hanle and inverted Hanle signal for a control device with structure
p-type Si/Al$_2$O$_3$/Au(10nm)/Ni$_{80}$Fe$_{20}$, in which
the non-magnetic Au interlayer causes the tunnel spin polarization
to be zero. Measurements are done at room temperature with a constant
current of -195$\mu$A (hole injection condition). A constant bias voltage
of about -172mV was subtracted from the data.}
\label{fig6}
\end{figure}

\indent Given that the experimental data deviates fundamentally from the theory, it is of the utmost importance to convincingly establish that the observed spin signals are genuine and originate from spin accumulation, rather than some kind of measurement artifact. Such potential artifacts can arise from (anisotropic) magnetoresistance effects related to the current through the ferromagnetic electrode itself, or from the effect of magnetic fields on charge transport in the semiconductor (Hall voltages etc.). A powerful way to exclude these artifacts is to introduce a thin non-magnetic layer at the interface between the tunnel oxide and the ferromagnet, without removing the ferromagnet \cite{patel}. The method relies on the extreme interface sensitivity of (spin-polarized) tunneling, such that insertion of a thin non-magnetic layer causes the tunnel spin polarization to vanish, and hence the spin accumulation. Genuine spin signals should then disappear, whereas any signals due to artifacts, if present, would still remain. This approach was previously used to rule out artifacts in the experiments by Dash et al. \cite{dash,dashspie}, although only the signal for out-of plane magnetic field (Hanle curve) was investigated, and only in the range of small field. In Fig. \ref{fig6}, a more complete characterization is presented, showing measurements on a control device in the Hanle as well as the inverted Hanle geometry, and for fields up to 50 kOe. No spin signals are observed. This implies that the signals (Hanle and inverted Hanle) observed in the regular devices (without the non-magnetic interlayer) are not due to an artifact but originate from spin-polarized tunneling and the spin accumulation this produces. This result corroborates previous experiments on spin injection from similar ferromagnetic tunnel contacts into a silicon light emitting diode \cite{jonker,jansencs}, from which the presence of spin-polarized carriers inside the silicon was unambiguously established.

\section{Discussion and conclusion}
\indent It has previously been noted that the magnitude of the spin accumulation signal observed
in magnetic tunnel devices on semiconductors is significantly different than that predicted
by the theory of spin injection, diffusion and detection, first for GaAs based devices \cite{tran},
and subsequently also for Si and Ge based devices \cite{jansennmatreview,dash}. The results
presented here provide an even larger discrepancy (of up to 7 orders of magnitude), and perhaps
more importantly, reveal that the scaling with tunnel barrier resistance deviates universally
from theory in a fundamental way. The scaling extends over a wide range of tunnel resistance,
down to the lowest tunnel RA values of about 10 k$\Omega\mu$m$^2$. It would certainly be of
interest to extend the measurements to devices with even lower tunnel resistance. Although
it was recently proposed that ferromagnetic tunnel contacts on Si with a single layer of
graphene as the tunnel barrier may be ideal for this purpose \cite{ervegraphene,derynandv},
the obtained tunnel RA product (6 k$\Omega\mu$m$^2$) was not much different from what was
already achievable with oxide tunnel contacts. For instance, in silicon-based non-local
devices, Fe/MgO contacts with a tunnel RA product of 4.6 k$\Omega\mu$m$^2$ have been
successfully used for spin injection and detection by Suzuki et al. \cite{suzuki}.
However, unlike the case of graphene, the oxide tunnel barriers can still be made
thinner and thus appear more promising to reach even lower RA product.\\
\indent Our results reveal that care has to be taken when in a particular device the observed magnitude of the spin signal is
found to be in agreement with theory, because this could be accidental. For instance, the scaling trend
predicts that at small junction RA product there must be a point where experiment and theory are in
agreement, but a more detailed investigation varying the tunnel barrier thickness would reveal a discrepancy.
This point of "accidental agreement"  will shift to larger junction RA product when the thickness of
the semiconductor channel is reduced, because the theory predicts a larger spin signal
when the volume of semiconductor into which spins are injected is decreased. An experiment to explicitly
confirm the predicted enhancement (for instance by studying devices with different channel thickness)
would be helpful, but is still lacking. Clearly, one needs to look beyond the magnitude of the
spin signal in order to (in)validate the theory.\\
\indent While the above results are obtained with three-terminal devices
and the observed signal is larger than predicted, in silicon-based non-local devices \cite{suzuki}
the observed signal deviates in the opposite direction, i.e., it is about two orders of magnitude
smaller than expected, as recently noted \cite{jansennmatreview}. Although in the latter
case there can be several other reasons, the results presented here suggest that the difference between
experiment and theory in three-terminal and non-local devices has a common origin, namely, a missing
ingredient in the existing theoretical descriptions. Several explanations for the discrepancy had
so far been proposed. These include two-step
tunneling via localized states near the semiconductor/oxide interface \cite{tran}, lateral inhomogeneity
of the tunnel current density \cite{dash}, two-step tunneling in parallel with direct tunneling \cite{jansentwostep},
or two-step tunneling via localized states near the oxide/ferromagnet interface \cite{uemura}. The scaling
results presented here, together with the control experiments, show unambiguously that none of
these proposals can explain the results. It is unclear whether the discrepancy arises from an
incorrect description of the magnitude of the spin accumulation that is induced by spin injection,
or from the description of the conversion of the induced spin accumulation into a voltage signal
in a Hanle measurement. Obviously, resolving this puzzle is of crucial importance for application
of magnetic tunnel contacts in semiconductor spintronic devices.\\

\section{Acknowledgements}
\indent The authors are grateful to T. Nozaki for his help with
the growth of the Ru-based control devices. The authors acknowledge
financial support from the Netherlands Foundation for Fundamental
Research on Matter (FOM), and from the Japanese Funding Program for
Next Generation World-Leading Researchers (No. GR099). One of the
authors (A.S.) acknowledges a JSPS Postdoctoral Fellowship for
Foreign Researchers.

\begin{appendix}
\section{Additional data}
\begin{figure}[htb]
\hspace*{0mm}\includegraphics*[width=60mm]{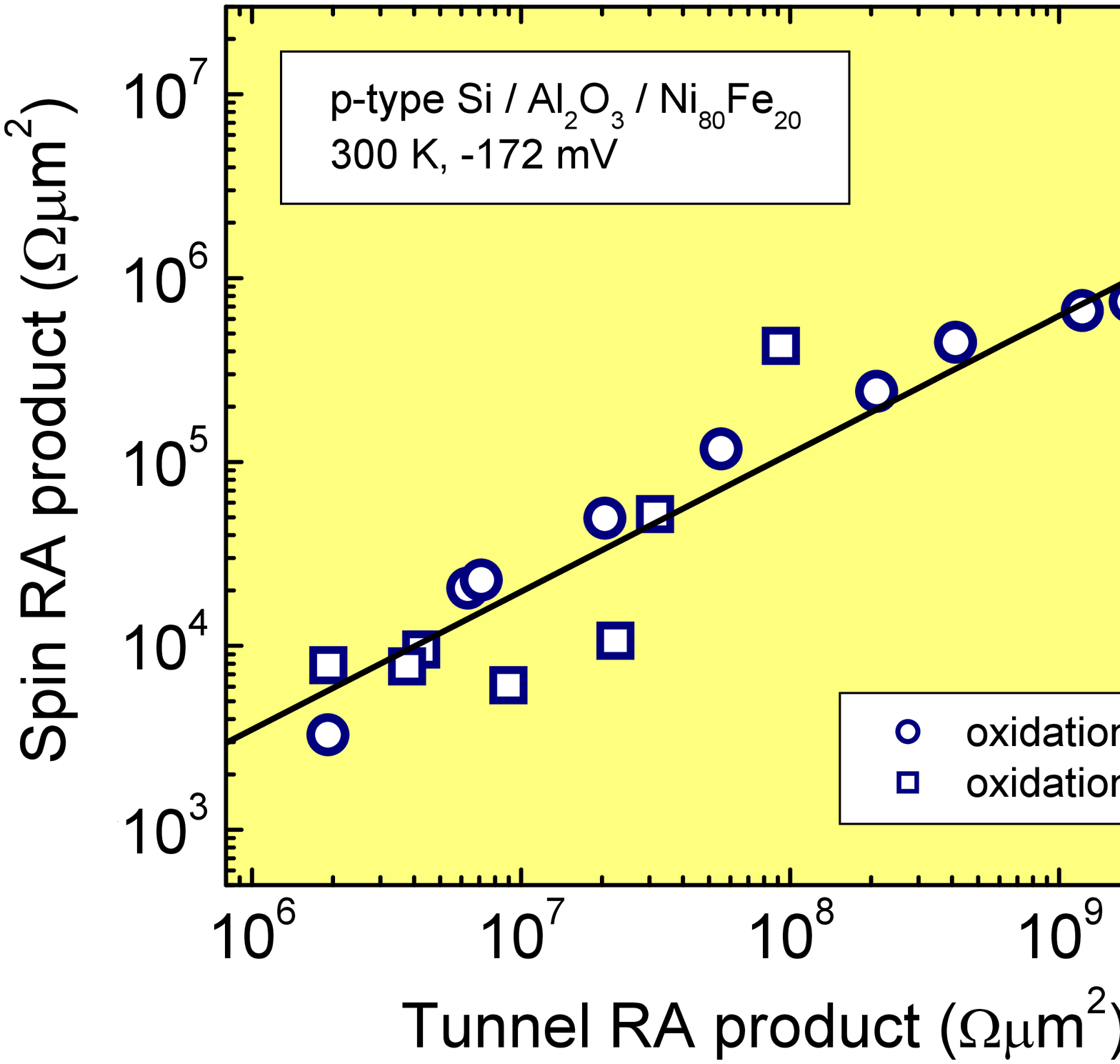}
\hspace*{0mm}\includegraphics*[width=59mm]{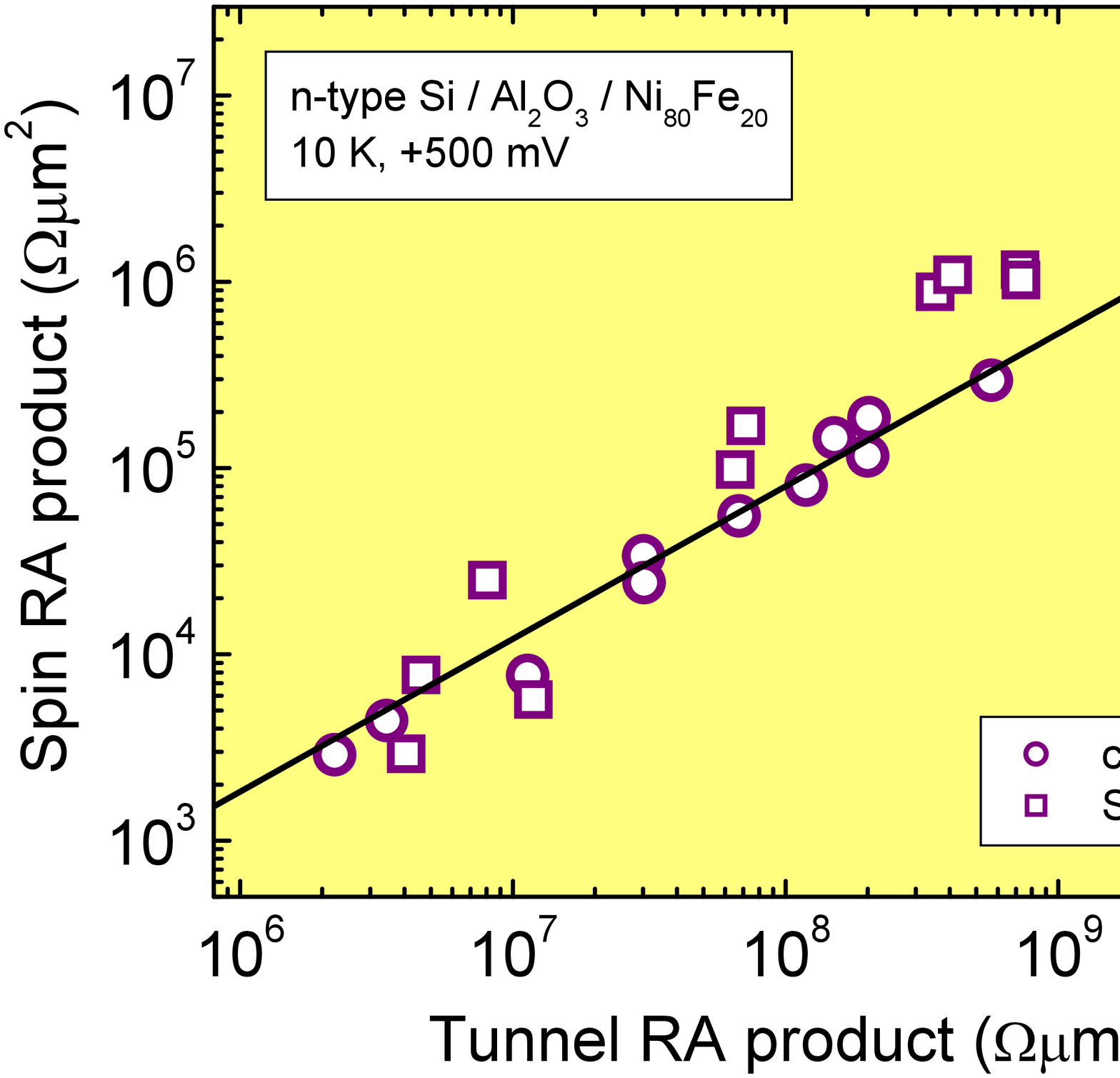}
\caption{Additional data on scaling of Hanle spin signals in
tunnel devices with n-type and p-type Si. The open circles are
the same data as in Fig. 2(b) and 2(c), whereas squares are additional data for devices with different
oxidation time (p-type, left panel) and with Cs treated surfaces (n-type,
right panel).}
\label{figS3}
\end{figure}

\indent To investigate the effect of localized states produced by oxygen vacancies within
the oxide tunnel barrier, we fabricated devices with p-type Si and
Al$_{2}$O$_{3}$ tunnel barrier, but without the plasma oxidation step. Since the
Al$_{2}$O$_{3}$ is grown by electron-beam deposition, the deposited oxide is oxygen deficient.
We found that there is no effect on the spin signal, i.e., junctions with and without
the plasma oxidation have the same spin RA product at the same tunnel resistance
(Fig. \ref{figS3}, left panel). This suggests that two-step tunneling via localized states
within the tunnel barrier plays no major role in the spin transport, consistent with the
result of the control devices with Ru metal.\\
\indent In order to investigate the effect of the resistance $r_b$ of the depletion region in the
Si, the Schottky barrier height was reduced (and with it $r_b$) using the procedure with a Cs
treatment of the Si surface that was previously developed \cite{dash,jansencs}. Here
we present similar data as in Ref. \onlinecite{dash} for n-type Si/Al$_{2}$O$_{3}$/Ni$_{80}$Fe$_{20}$
devices with and without Cs, but now as a function of tunnel barrier thickness
(Fig. \ref{figS3}, right panel). The Cs treatment produces no change of the spin RA product, and
it scales to values of 10$^6$ $\Omega \mu$m$^2$. This is not compatible with a description
in terms of two-step tunneling via localized states at the oxide/Si interface. Owing to the
small value of $r_b$ for the devices treated with Cs, a large spin accumulation cannot built up
in the interface states because spins will leak away efficiently into the silicon.

\section{Hanle line width versus tunnel barrier thickness}
\begin{figure}[h]
\hspace*{0mm}\includegraphics*[width=68mm]{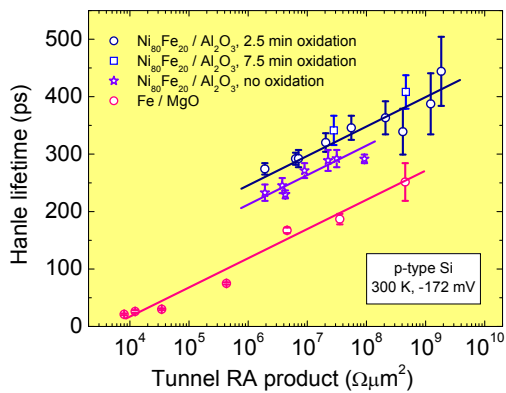}
\caption{Hanle line width versus tunnel resistance. The Hanle line width is characterized by an effective lifetime obtained from a fit of the Hanle curve using a Lorentzian. This time constant is a lower bound to the spin lifetime, as previously discussed \cite{dash,invertedhanle,jansensstreview}. The solid lines are guides to the eye.}
\label{figS5}
\end{figure}

\indent Fig. \ref{figS5} shows that the effective spin lifetime, extracted from the width of the Hanle curves, increases as a function of the tunnel resistance. For devices with an Al$_{2}$O$_{3}$ tunnel barrier, the line width is slightly dependent on the oxidation time. The spin lifetime for devices with MgO/Fe contacts is smaller than with Al$_{2}$O$_{3}$/Ni$_{80}$Fe$_{20}$ contacts. This is attributed to broadening of the Hanle curve by inhomogeneous magnetostatic fields, which is more pronounced for Fe owing to its larger magnetization \cite{invertedhanle}.

\end{appendix}



\begin{thebibliography}{99}

\footnotesize

\bibitem{chappert} C. Chappert, A. Fert, and F. Nguyen van Dau, Nature
Mater. {\bf 6}, 813 (2007).

\bibitem{awschalom} D. D. Awschalom and M. E. Flatt\'{e}, Nature Phys. {\bf 3}, 153 (2007).

\bibitem{fertnobel} A. Fert, Rev. Mod. Phys. {\bf 80}, 1517 (2008).

\bibitem{jansennmatreview} R. Jansen, Nature Mater. {\bf 11}, 400 (2012).

\bibitem{jonker} B. T. Jonker, G. Kioseoglou, A. T. Hanbicki, C. H. Li, and P. E. Thompson, Nature Phys. {\bf 3}, 542 (2007).

\bibitem{erve} O. M. J. van 't Erve {\em et al.}, Appl. Phys. Lett. {\bf 91},
212109 (2007).

\bibitem{dash} S. P. Dash, S. Sharma, R. S. Patel, M. P. de Jong, and R. Jansen, Nature {\bf 462}, 491 (2009).


\bibitem{jansencs} R. Jansen, B. C. Min, S. P. Dash, S. Sharma, G. Kioseoglou, A. T. Hanbicki,
O. M. J. van 't Erve, P. E. Thompson, and B. T. Jonker, Phys. Rev. B {\bf 82}, 241305 (2010).

\bibitem{sasakitdep} T. Sasaki, T. Oikawa, T. Suzuki, M. Shiraishi, Y. Suzuki, and K. Noguchi, Appl. Phys. Lett. {\bf 96},
122101 (2010).

\bibitem{suzuki} T. Suzuki, T. Sasaki, T. Oikawa, M. Shiraishi, Y. Suzuki, K. Noguchi, Appl. Phys. Express {\bf 4}, 023003 (2011).

\bibitem{jeon} K. R. Jeon, B. C. Min, I. J. Shin, C. Y. Park, H. S. Lee, Y. H. Jo, and S. Ch. Shin, Appl. Phys. Lett. {\bf
98}, 262102 (2011).

\bibitem{hamayasi3} Y. Ando, K. Kasahara, K. Yamane, Y. Baba, Y. Maeda, Y. Hoshi, K. Sawano, M. Miyao,
and K. Hamaya, Appl. Phys. Lett. {99}, 012113 (2011).

\bibitem{hamayaefield} Y. Ando, Y. Maeda, K. Kasahara, S. Yamada, K. Masaki, Y. Hoshi, K. Sawano,
K. Izunome, A. Sakai, M. Miyao, and K. Hamaya, Appl. Phys. Lett. {\bf 99}, 132511 (2011).

\bibitem{toshiba1} T. Inokuchi, M. Ishikawa, H. Sugiyama, Y. Saito, and N. Tezuka, J. Appl. Phys. {\bf 111}, 07C316 (2012).

\bibitem{toshiba2} M. Ishikawa, H. Sugiyama, T. Inokuchi, K. Hamaya, and Y. Saito, Appl. Phys. Lett. {\bf 100}, 252404 (2012).

\bibitem{saitoge} H. Saito, S. Watanabe, Y. Mineno, S. Sharma, R. Jansen, S. Yuasa, and K. Ando, Solid State Comm. {\bf 151}, 1159 (2011).

\bibitem{wangge} Y. Zhou, W. Han, L. T. Chang, F. Xiu, M. Wang, M. Oehme, I. A. Fischer,
J. Schulze, R. K. Kawakami, and K. L. Wang, Phys. Rev. B {\bf 84}, 125323 (2011).

\bibitem{jeonge} K. R. Jeon, B. C. Min. Y. H. Jo, H. S. Lee, I. J. Shin,
C. Y. Park, S. Y. Park, and S. Ch. Shin, Phys. Rev. B {\bf 84}, 165315 (2011).

\bibitem{jain} A. Jain {\em et al.}, Appl. Phys. Lett. {\bf 99}, 162102 (2011).

\bibitem{ibage} S. Iba, H. Saito, A. Spiesser, S. Watanabe, R. Jansen, S. Yuasa,
and K. Ando, Appl. Phys. Express {\bf 5}, 023003 (2012).

\bibitem{ibagert} S. Iba, H. Saito, A. Spiesser, S. Watanabe, R. Jansen, S. Yuasa,
and K. Ando, Appl. Phys. Express {\bf 5}, 053004 (2012).

\bibitem{hamayaschottky} K. Kasahara, Y. Baba, K. Yamane, Y. Ando, S. Yamada,
Y. Hoshi, K. Sawano, M. Miyao, and K. Hamaya, J. Appl. Phys. {\bf 111},
07C503 (2012).

\bibitem{jansensstreview} R. Jansen, S. P. Dash, S. Sharma, and B. C. Min,
Semicond. Sci. Technol. {\bf 27}, 083001 (2012).

\bibitem{fertPRB} A. Fert and H. Jaffr\`{e}s, Phys. Rev. B {\bf 64}, 184420 (2001).


\bibitem{fertIEEE} A. Fert, J. -M. George, H. Jaffr\`{e}s, and R. Mattana, IEEE
Trans. Elec. Dev. {\bf 54}, 921 (2007).

\bibitem{maekawa} S. Takahashi and S. Maekawa, Phys. Rev. B {\bf 67}, 052409 (2003).

\bibitem{dery} Y. Song and H. Dery, Phys. Rev. B {\bf 81}, 045321 (2010).

\bibitem{tran} M. Tran, H. Jaffr\`es, C. Deranlot, J. -M. George, A. Fert, A. Miard,
and A. Lema\^itre, Phys. Rev. Lett. {\bf 102}, 036601 (2009).

\bibitem{jansentwostep} R. Jansen, A. M. Deac, H. Saito, and S. Yuasa, Phys. Rev. B {\bf 85}, 134420 (2012).

\bibitem{minthesis} B. C. Min, Ph.D. Thesis (Koninklijke W\"{o}hrmann, Zutphen, The Netherlands, 2007). pp. 19-22.

\bibitem{spiesserspie} A. Spiesser, S. Sharma, H. Saito, R. Jansen, S. Yuasa, and K. Ando, Proc. SPIE {\bf 8461}, 84610K
(2012).

\bibitem{invertedhanle} S. P. Dash, S. Sharma, J. C. Le Breton, J. Peiro, H. Jaffr\`{e}s, J. -M.
George, A. Lema\^{i}tre, and R. Jansen, Phys. Rev. B {\bf 84}, 054410
(2011).

\bibitem{robertson} J. Robertson, Rep. Prog. Phys. {\bf 69}, 327 (2006).

\bibitem{beasley} Y. Xu, D. Ephron, and M. R. Beasley, Phys. Rev. B {\bf 52}, 2843 (1995).

\bibitem{uemura} T. Uemura, K. Kondo, J. Fujisawa, K. Matsuda, and M. Yamamoto, Appl. Phys. Lett. {\bf 101}, 132411 (2012).

\bibitem{valenzuela} S. O. Valenzuela, D. J. Monsma, C. M. Marcus, V. Narayanamurti and M. Tinkham,
Phys. Rev. Lett. {\bf 94}, 196601 (2005).

\bibitem{park} B. G. Park, T. Banerjee, J. C. Lodder and R. Jansen, Phys. Rev. Lett. {\bf 99}, 217206 (2007).

\bibitem{fabianacta} J. Fabian, A. Matos-Abiague, C. Ertler, P. Stano, and I. \v{Z}uti\'{c},
Acta Physica Slovaca {\bf 57}, 565 (2007).


\bibitem{yuasa} S. Yuasa and D. D. Djayaprawira, J. Phys. D: Appl. Phys. {\bf 40}, R337 (2007).

\bibitem{patel} R. S. Patel, S. P. Dash, M. P. de Jong, and R. Jansen, J. Appl. Phys. {\bf 106}, 016107 (2009).

\bibitem{dashspie} S. P. Dash, S. Sharma, J. C. Le Breton, and R. Jansen, Proc. SPIE {\bf 7760}, 77600J (2010).

\bibitem{ervegraphene} O. M. J. van 't Erve, A. L. Friedman, E. Cobas, C. H. Li, J. T. Robinson, and B. T. Jonker,
Nature Nanotech. {\bf 7}, 737 (2012).

\bibitem{derynandv} H. Dery, Nature Nanotech. {\bf 7}, 692 (2012).

\end{thebibliography}
\end{document}